\documentclass[preprint]{iucr}              
\usepackage{graphicx}
\usepackage{amssymb}
\usepackage{url}
\usepackage{bm}
\usepackage[fleqn]{amsmath}
\usepackage[yyyymmdd,hhmmss]{datetime}

\newcommand{\vdotv}[2]{${\mathbf{#1} \cdot \mathbf{#2}}$}

     \journalcode{A}              

\begin{document}                  



\title{An Invertible Seven-Dimensional Dirichlet Cell Characterization of Lattices}
\shorttitle{DC7unsrt}


\cauthor[a]{Herbert J.}{Bernstein}{yayahjb@gmail.com}{}
\author[b]{Lawrence C.}{Andrews}
\author[c,d]{Mario}{Xerri}

\aff[a]{Ronin Institute for Independent Scholarship, c/o NSLS-II, Brookhaven National Laboratory, Bldg 745, P.O. Box 5000, Upton, NY \country{USA}}
\aff[b]{Ronin Institute for Independent Scholarship, Kirkland, WA \country{USA} }
\aff[c]{College of Engineering, Cornell Univ., Ithaca, NY \country{USA} }
\aff[d]{Brookhaven National Laboratory, Bldg 745, P.O. Box 5000, Upton, NY \country{USA}}

\shortauthor{Bernstein, Andrews and Xerri}







\maketitle                        

\begin{synopsis}
Starting from a Niggli-reduced cell, a crystallographic lattice may be characterized by seven parameters describing the Dirichlet cell: the three 
shortest non-coplanar lattice vector lengths, the three shorter 
of each pair of face diagonal lengths and the shortest body diagonal length,
from which the Niggli-reduced cell may be recovered.
\end{synopsis}

\begin{abstract}
Characterization of crystallographic lattices is an important tool in
structure solution, crystallographic database searches and
clustering of diffraction images in serial crystallography.
Characterization of lattices by Niggli-reduced cells
(based on the three shortest non-coplanar lattice vectors)
or by Delaunay-reduced cells (based on four non-coplanar 
vectors summing to zero
and all meeting at obtuse or right angles) are commonly used.  The
Niggli cell derives from Minkowski reduction.  The Delaunay cell derives
from Selling reduction.  All are related to the Wigner-Seitz (or
Dirichlet, or Voronoi) cell of the lattice, which consists of the
points at least as close to a chosen lattice point as they are to any 
other lattice point.  We call the three non-coplanar lattice vectors 
chosen the Niggli-reduced cell edges.
Starting from a Niggli-reduced cell, the Dirichlet cell is
characterized by the planes determined by thirteen lattice half-edges:
the midpoints of the three Niggli cell edges, the six Niggli cell face-diagonals and the four body-diagonals, but seven of the
lengths are sufficient:  three edge lengths, the three shorter of
each pair of face-diagonal lengths, and the shortest body-diagonal length.
These seven are sufficient to recover the 
Niggli-reduced cell.

\end{abstract}


\section{Introduction}

Algorithms for quantifying the differences among lattices are used for 
Bravais lattice determination, for database lookup of unit cells to select 
candidates for molecular replacement, and for clustering to 
group together images from serial crystallography.
In order to 
create a distance measure, it is necessary to define a metric
representation of lattices. The present paper describes a new
representation with sufficient detail for creating a complete
distance measure.

 For crystallography,
there are many alternative representations to choose from as 
a basis for distance calculations. 

$\bullet$ \citeasnoun{Andrews1980} defined
$\mathbf{V}^7$, a perturbation-stable space in which, using real and 
reciprocal space Niggli
reduction, a lattice is represented by three cell-edge lengths, three reciprocal 
cell-edge lengths, and the cell volume.  While suitable for database
searches, $\mathbf{V}^7$ was found to have difficulties in some other uses. For lattice type determination, issues arose when working near right angles.
 
$\bullet$ \citeasnoun{Andrews1988} then
defined $\mathbf{G}^6$ that uses a modified metric tensor and an iterative search through 25 alternative
reduction boundary transforms \cite{Gruber1973} to work in a satisfactory manner both for
database searches and lattice identification in the presence of experimental
error.

$\bullet$ \citeasnoun{Andrews2014} discussed sewing together regions of the fundamental
region of $\mathbf{G}^6$ under Niggli reduction at fifteen boundaries.

$\bullet$ \citeasnoun{Andrews2019}
presented the simplest and  fastest currently known representation of lattices as the six Selling 
scalars obtained from the dot products of the unit cell axes in 
addition to the negative of their sum (a body diagonal). Labeling 
three linearly independent vectors $\mathbf{a},\, \mathbf{b}, \mathbf{c}$ spanning the lattice, and defining $\mathbf{d}$ ($\mathbf{d} = -\mathbf{a}-\!\mathbf{b}-\!\mathbf{c}$), 
the Selling scalars are
	
\begin{center}
	\vdotv{b}{c},~ \vdotv{a}{c},~ \vdotv{a}{b},
	~ \vdotv{a}{d},~ \vdotv{b}{d},~ \vdotv{c}{d}
\end{center}

\noindent{}(where, {\it e.g.}, \vdotv{b}{c}   represents the dot product of 
	the $\bf{b}$  and $\bf{c}$ vectors). For the purpose of organizing 
	these six quantities as a vector space in which one can
compute simple Euclidean distances, we describe the set of scalars as a 
vector, $\bf{s}$, with components, $ s_1, s_2, s_3, ..., s_6 $.  
The cell is Selling-reduced if all six components are negative or zero \cite{Delone1933}.  Reversing Allman's observation that a Buerger-reduced 
cell is a good stepping stone to a Selling-reduced cell \cite{Allmann1968},
a Selling-reduced cell can be an efficient stepping stone to
a  Buerger-reduced cell that quickly reduces to a Niggli-reduced cell \cite{Andrews2019}.
Note that the negatives of the Selling scalars of a Selling-reduced cell
are non-negative so that the six square roots provide a convenient six-parameter 
characterization of a lattice \cite{kurlin2022complete}.

\subsection{A new representation}

In this paper we consider lattice representation based on
the Wigner-Seitz \cite{wigner1933} (or Dirichlet, or Voronoi) cell of the lattice, which consists of the points at least as close to a chosen lattice point as they are to any other lattice point.  
Starting from a Niggli-reduced \cite{Niggli1928} cell, the Dirichlet cell is
characterized by the planes determined by thirteen lattice half-edges:
the midpoints of the three Niggli-reduced cell edges, the six Niggli-reduced cell face-diagonals, and the four body-diagonals, but seven of the Niggli-reduced cell lattice vector
lengths are sufficient:  three Niggli-reduced cell-edge lengths, the three shorter of each pair of 
face-diagonal lengths, and the shortest body-diagonal length.
The Niggli-reduced cell may be recovered easily from these seven
quantities.

\begin{figure}
\caption{Truncated octahedron  \cite{wikipeditrancatedoctahedron}.
Image licensed under the Creative Commons Attribution-Share Alike 3.0 Unported license. Subject to disclaimers.  See web site.
}
\label{fig:truncoctahedron}
\begin{center}
\includegraphics[scale=.5]{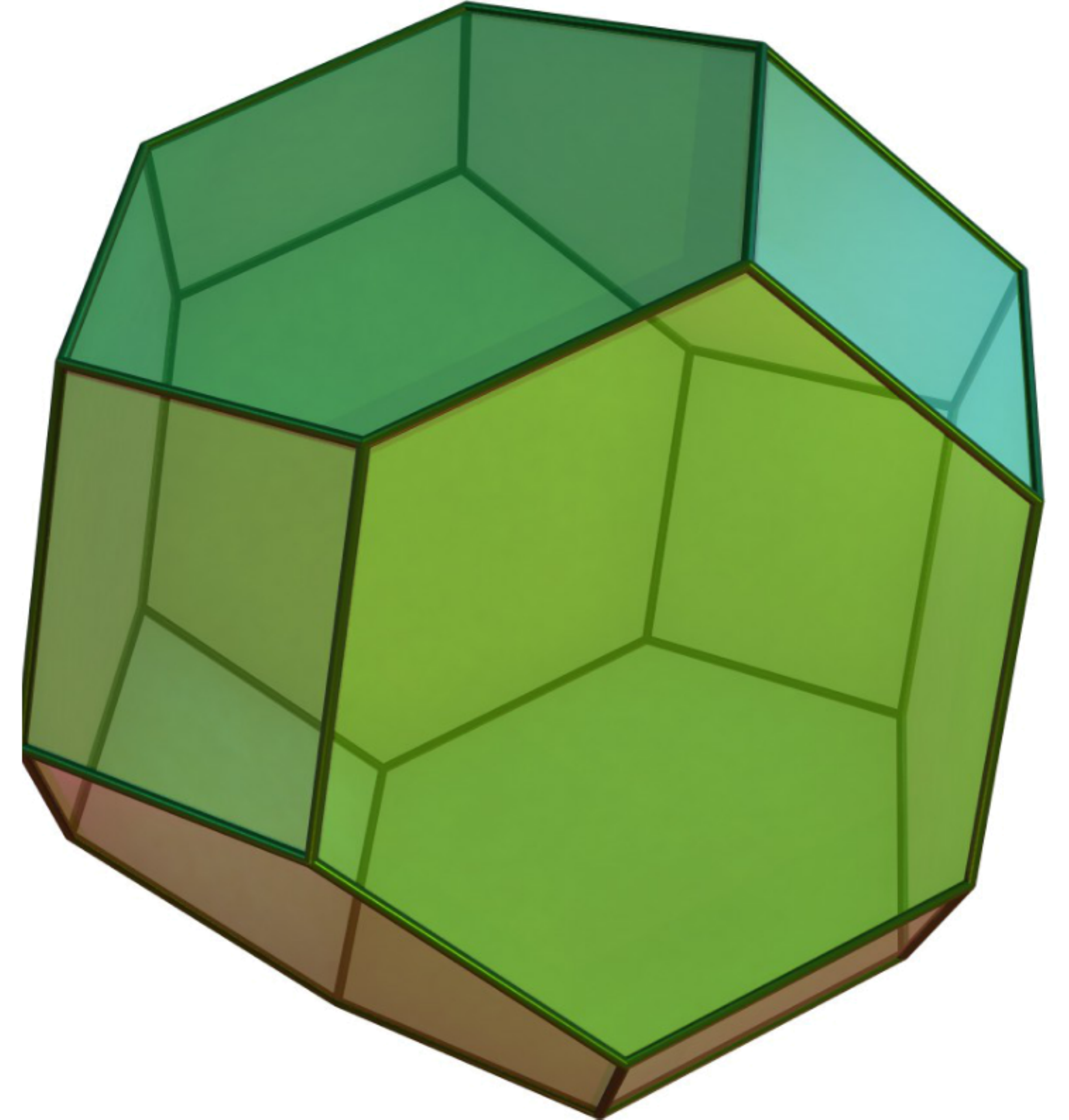}
\end{center}
\end{figure}

A Wigner-Seitz cell is a polyhedron of six, eight, ten, twelve or fourteen
faces.  The general fourteen-face case is a truncated octahedron.  See
Fig. \ref{fig:truncoctahedron}.

When one is creating a metric to compute distances between lattices the metric
is likely to be based directly or indirectly either on the components
of the Niggli-reduced cell or on the components of the Selling-reduced
cell.  One disadvantage in use of the former is that the space
of Niggli-reduced cells admits both all-acute and all-obtuse presentations,
dividing the space into two distinct components with a disruptive boundary
between them.  Use of Selling-reduction avoids this problem by restricting
one's attention to the all-obtuse case, making it easier to find smooth
metrics.  The Wigner-Seitz cell is often presented entirely in the context
of Selling-reduction, but in this paper we are using the approach of \cite{Hart2019}
and working in terms of Niggli-reduction in order to understand the primary
static characteristics of the space, especially the invertibility of
this presentation.  The question of improving stability, especially when
computing distances between lattices in the $---$ and $+++$ components, is left for future
investigation.

\section{Background}

\begin{figure}
\caption{Historical timeline of studies of crystallographic lattice
characterization.  Figure drawn by E. Kincaid.  Used with
permission of the artist.
}
\label{fig:timeline}
\begin{center}
\includegraphics[scale=.5]{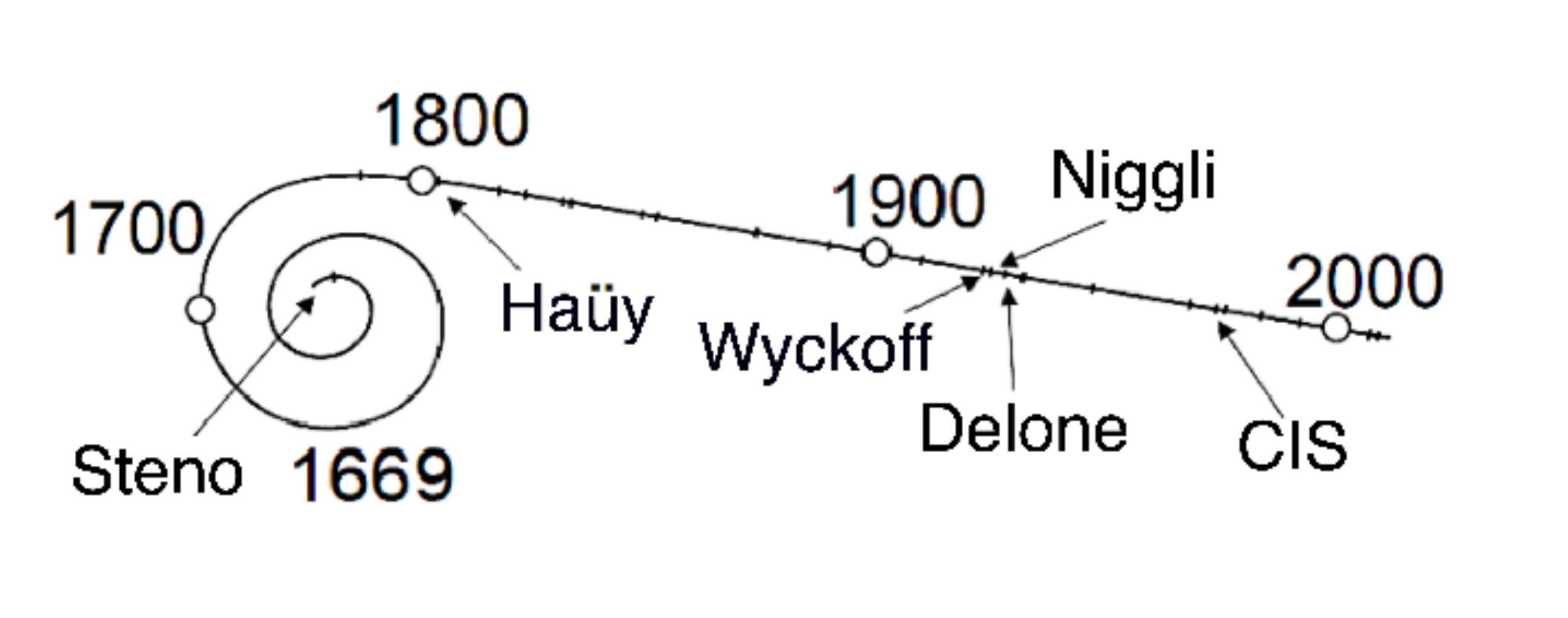}
\end{center}
\end{figure}

That crystals are built from some regular assembly of basic parts was already clear in ancient times.    In 1611 Kepler described this relationship (\citeasnoun{Kepler1611}, translated in \citeasnoun {kepler1966}). Steno was asked to prepare a catalog of a ``cabinet of curiosities''; this is considered the first database of crystals (minerals in this case) \cite{Steno1669}.  See Fig. \ref{fig:timeline} for a timeline
of lattice characterization from Steno onwards.  In the 19th century 
indices were published with interaxial angles of crystals, specifically 
for the identification of minerals.

Following the discovery of x-rays, 
catalogs of unit cell parameters started to be published \cite{wyckoff1931}.
Often, these were arranged by crystal system and then sorted by some of the cell parameters. However, related minerals with distortions or deformed into another crystal system or incorrectly attributed to another could be difficult to find. Clearly, a metric for relating unit cells was required.  See \citeasnoun{Andrews2023} for additional
background information on the problem.

\subsection{Reduced Cells}

\citeasnoun{Niggli1928} and \citeasnoun{Delaunay1933} (aka Delone) devised ``reduced cells'', which allowed for a more standard presentation of some of the crystal data.
The Buerger-reduced cell is simpler than the Niggli-reduced cell, having fewer
constraints \cite{Buerger1957, Azaroff1958, Buerger1960}.  All Niggli-reduced cells are Buerger-reduced, but not all Buerger-reduced cells are Niggli-reduced.  Roof \cite{Roof1967} provided an updated and authoritative reference to the lattice characters of Niggli-reduced cells.

Finally, in the 1970s, the National Institute of Health (NIH) and the Environmental Protection Agency (EPA) joined to create the online searchable Chemical Information System (CIS) \cite{Heller1976, Bernstein1979}.
Along with physical measurements such as nuclear magnetic resonance and infrared, NIH/EPA wanted to include unit cell searching. At the time, there was no commonly accepted method to compute the ``distance'' between two unit cells (equivalently, lattices).  

There were two problems.
	
The first problem was that measured unit cell parameters (conventionally, [a, b, c, alpha, beta, gamma]  for the cell lengths and angles) always have experimental error in their determinations.  Closely related compounds of interest might have slightly different cell-edge parameters.  That means that the problem to be solved is ``the nearest neighbor problem'' also known as ``the post office problem''. An exact match is inadequate to find
closely related but non-identical neighbors.

The second issue is related to the problem of experimental error, but it manifests in a different way. It is well-known that for any given lattice, there is an infinity of unit cells that can be chosen.  The problem is that two unit cells from the same lattice  may not look the same.

\section{The Unsorted $\mathbf{DC}^7$ Cell, dc7unsrt}

We define the Wigner-Seitz cell as consisting of the points which
are no farther from a given lattice point than they are from any
other lattice point.   As \citeasnoun{Hart2019} has shown, the
Wigner-Seitz cell centered on a given lattice point is contained 
entirely within the convex envelope of the immediate 
neighbors of a given lattice point, {\it i.e.} in terms of twenty-six
Miller indices:
~~\\
\vspace{-5mm}
\[
(1,0,0),\,(0,1,0),\,(0,0,1),\,(-1,0,0),\,(0,-1,0),\,(0,0,-1),
\]
~~\\

\vspace{-30mm}
\[(0,1,1),\,(1,0,1),\,(1,1,0),\,(0,1,-1),\,(-1,0,1),\,(1,-1,0),
\]
~~\\

\vspace{-30mm}
\[
(0,-1,\,-1),\,(-1,0,-1),\,(-1,-1,0),\,(0,-1,1),\,(1,0,-1),\,(-1,1,0),
\]
~~\\

\vspace{-30mm}
\[
(1,1,1),(1,-1,-1),\,(-1,1,-1),\,(-1,-1,1),
\]
~~\\

\vspace{-30mm}
\[
(-1,-1,-1),(-1,1,1),(1,-1,1,),(1,1,-1)
\]

We organize the lattice in terms of a basis of the three shortest distances.
The Wigner-Seitz cell is symmetric around the given lattice point, so thirteen Miller indices,
~~\\

\vspace{-20mm}
\[
(1,0,0),\,(0,1,0),\,(0,0,1),
\]
~~\\

\vspace{-30mm}
\[
(0,1,1),\,(1,0,1),\,(1,1,0),\,(0,1,-1),\,(-1,0,1),\,(1,-1,0),
\]
~~\\

\vspace{-30mm}
\[
(1,1,1),\,(1,-1,-1),\,(-1,1,-1),\,(-1,-1,1)
\]

\noindent{}are sufficient.

Formally, the definition of the Wigner-Seitz cell is:

Let $\mathbf{R}$ be the space of reals, $\mathbf{Z}$ be the
space of integers.  Let $L$ be an $\mathbf{R}^3$ lattice with Minkowski basis $\mathbf{a},\, \mathbf{b},\, \mathbf{c}\, \in \mathbf{R}^3$, {\it i.e.} such that $h\mathbf{a}+k\mathbf{b}+l\mathbf{c}, h \in \mathbf{Z}, k \in \mathbf{Z}, l \in \mathbf{Z}$ spans $L$ and $||\mathbf{a}||, ||\mathbf{b}||, ||\mathbf{c}||$ are minimal.  We define the Wigner-Seitz cell of $L$ as
\[
WS(L) = \{ w \in \mathbf{R}^3 \backepsilon \forall x \in L, x \neq 0, ||w|| \leq ||x-w|| \}
\] 

If we translate this cell to each element of $L$, we tile the space and have a
Voronoi decomposition.

Niggli-reduction provides an unambiguous Minkowski reduction. Assume the cell formed by $\mathbf{a}, \mathbf{b}, \mathbf{c}$ is Niggli-reduced, with  $||\mathbf{a}||\, \leq\, ||\mathbf{b}|| \leq ||\mathbf{c}||$.  Define the $\mathbf{G}^6$ vector of
the cell as
\[
\{r,\,s,\,t,\,u,\,v,\,w\}
\]
\[=\{ \mathbf{a} \cdot \mathbf{a},\, \mathbf{b} \cdot \mathbf{b},\, \mathbf{c} \cdot \mathbf{c},\,  2 \mathbf{b} \cdot \mathbf{c},\, 2 \mathbf{a} \cdot \mathbf{c},\, 2 \mathbf{a} \cdot \mathbf{b}\}
\]
As a Niggli-reduced cell, we either have all of $u,v,w \leq 0$, which we
annotate as the $---$ case, or all of $u,v,w > 0$, which we annotate
as the $+++$ case, and in both cases $|u| \leq s,\, |v| \leq r,\, |w| \leq r$.

The full set of conditions for a Niggli-reduced cell are
given in conditions \ref{eq:niggli_start} -- \ref{eq:niggli_end}:
\begin{equation}
\text{require } 0 < r \leq s \leq t \label{eq:niggli_start}
\end{equation}
\begin{equation}
\text{if } r = s; \text{ then require } |u| \leq |v|
\end{equation}
\begin{equation}
\text{if } s = t; \text{ then require } |v| \leq |w| 
\end{equation}
\begin{equation}
\text{require } u > 0 \text{ and } v > 0 \text{ and } w > 0 \text{ or}
\end{equation}
\begin{equation}
\text{~~~~require } u \leq 0 \text{ and } v \leq 0 \text{ and } w \leq 0 
\end{equation}
\begin{equation}
\text{require } |u| \leq s
\end{equation}
\begin{equation}
\text{require } |v| \leq r
\end{equation}
\begin{equation}
\text{require } |w| \leq r
\end{equation}
\begin{equation}
\text{require } t \leq r + s + t + u + v + w 
\end{equation}
\begin{equation}
\text{if } u = s \text{ then require }  w \leq 2\,v
\end{equation}
\begin{equation}
\text{if } v = r \text{ then require }  w \leq 2\,u
\end{equation}
\begin{equation}
\text{if } w = r \text{ then require }  v \leq 2\,u
\end{equation}
\begin{equation}
\text{if } u = -s \text{ then require }  w = 0
\end{equation}
\begin{equation}
\text{if } v = -r \text{ then require }  w = 0
\end{equation}
\begin{equation}
\text{if } w = -r \text{ then require }  v = 0
\end{equation}
\begin{equation}
\text{if } t = r + s + t + u + v + w 
\text{ then require } 2\,r + 2\, v + w \leq 0 \label{eq:niggli_end}
\end{equation}

Using the $\mathbf{G}^6$ vector component definitions, we define a  $\mathbf{DC}^{13}$ cell as the squares of the three edge lengths, the six face diagonal lengths and the
four body diagonal lengths, {\it i.e.}:
\[
dc_{13,raw}(r,s,t,u,v,w)=
\]
\[
[r,\,s,\,t,
\]
\[
s+t-u,\, s+t+u,\, r+t-v,\, r+t+v,\, r+s-w,\, r+s+w, 
\]
\[
r+s+t+u+v+w,\, r+s+t+u-v-w,
\]
\[
r+s+t-u+v-w,\, r+s+t-u-v+w]
\]

If we sort the elements of $\mathbf{DC}^{13}$ and only present the first
seven elements, we have $\mathbf{DC}^{7}$ as discussed in \cite{bernstein2021},
which is a smooth, but ambiguous, characterization of lattices.  $\mathbf{DC}^{7}$ 
is not invertible in some cases unless the symmetry is known 
{\it a priori}, or some elements after the seventh are retained.  Bright
\cite{bright2021a} has demonstrated the $\mathbf{DC}^{7}$ ambiguity with
the cells i: $ [ 2.8284,$ $ 3.162277$, $ 3.4641 $, $ 117.157$, $ 107.8295$, $116.5651 ] $
and ii: $ [ 2.8284$, $ 3.162277$, $ 3.4641$, $ 123.211$, $ 107.8295$, $ 109.59748 ] $
as $ [a,b,c,\alpha,\beta,\gamma] $, or
i: $ [8,\,10,\,12,\,-10,\,-6,\,-4]$ and ii: $ [8,\,10,\,12,\,-12,\,-6,\,-3 ] $
as the $\mathbf{G}^6$ vectors $[r,\,s,\,t,\,u,\,v,\,w]$
for which the sorted $\mathbf{DC}^{13}$ elements after Niggli reduction are
~~\\

\vspace{-20mm}
\[ \text{i:}\, 2.44949,\, 2.82843,\, 3.16228,\]
~~\\

\vspace{-30mm}
\[ 3.16228,\, 3.4641,\, 3.4641,\, 3.74166,\,\mathbf{4},\]
~~\\

\vspace{-30mm}
\[4.24264,\, 4.47214,\, 5.09902,\,  5.09902,\, 6.16441, \]
~~\\

\vspace{-20mm}
\noindent{}and 
~~\\

\vspace{-20mm}
\[ \text{ii:}\, 2.44949,\, 2.82843,\, 3.16228,\]
~~\\

\vspace{-30mm}
\[ 3.16228,\, 3.4641,\, 3.4641,\, 3.74166,\, \mathbf{4.24264},\]
~~\\

\vspace{-30mm}
\[4.24264,\, 4.89898,\, 4.89898,\, 5.2915,\, 5.65685, \]
~~\\

\vspace{-20mm}
\noindent{}respectively, which do not differ until the eighth element.

If all of $u,\,v,\,w \leq 0$, then $s+t+u \leq s+t-u$, 
$r+t+v \leq r+t-v$, $r+s+w \leq r+s-w$, {\it i.e.} the three summed
squares of the face diagonals are no larger than the corresponding differences,
and $r+s+t+u+v+w \leq $ \{ $r+s+t+u-v-w$, $r+s+t-u+v-w$, $r+s+t-u-v+w$ \}.
In this $---$ case, the main body diagonal is no larger than the three remaining body diagonals.

On the other hand, if all of $u,v,w > 0$, then $s+t+u > s+t-u$, 
$r+t+v > r+t-v$, $r+s+w > r+s-w$; {\it i.e.} the squares of the
lengths of longer face diagonals formed by the Niggli-reduced
cell-edge vector sums are strictly greater than the corresponding
shorter face diagonals formed by the Niggli-reduced
cell-edge vector differences.  Turning to the body diagonals,
the square of the length of the main body diagonal is
$r+s+t+u+v+w$.  The squares of the lengths of the other three
body diagonals are $ \{r+s+t+u-v-w$, $r+s+t-u+v-w$, $r+s+t-u-v+w\}$.
Thus the square of the length of the smallest squared body diagonal is strictly less than $r+s+t+u+v+w$, and strictly greater than $r+s+t-|u|-|v|-|w|$.

Thus we can recover $r,s,t,u,v,w$ from the three cell-edge lengths,
the three shorter of each pair of face diagonal lengths, and the shortest squared body diagonal length
as explained in the following theorem and proof.

\subsection{Theorem}
\label{subsec::invert}

Let $g = {r,s,t,u,v,w}$ be a $\mathbf{G}^6$ Niggli-reduced
cell satisfying the Niggli-reduction conditions 
\ref{eq:niggli_start} -- \ref{eq:niggli_end}.  Let
\[ dc7_{unsrt}=[dc7_{unsrt,1},dc7_{unsrt,2},dc7_{unsrt,3},\]
\[ dc7_{unsrt,4},dc7_{unsrt,5},dc7_{unsrt,6},dc7_{unsrt,7} ] \]
\[ =[r,s,t,s+t-|u|,r+t-|v|,r+s-|w|,\] 
\[ min(r+s+t+u+v+w, r+s+t+u-v-w,\]
\[ r+s+t-u+v-w, r+s+t-u-v+w)]
\]
\[ =[r,s,t,s+t-|u|,r+t-|v|,r+s-|w|,\] 
\[ r+s+t-|u|-|v|-|w|+2\,min(max(0,u),max(0,v),max(0,w))]
\]
be the unsorted $\mathbf{DC}^{7}$ presentation of $g$.
Then the values of the components of $dc7_{unsrt}$
unambiguously determine the values of the components
of $g$.

\subsubsection{Proof of Theorem \ref{subsec::invert}}
~~\\

In the Niggli-reduced $---$ case we have
\begin{equation}
  dc7_{unsrt} = [r,s,t,s+t+u,r+t+v,r+s+w,r+s+t+u+v+w]
  \label{mmm}
\end{equation}
and in the Niggli-reduced $+++$ case we have
\begin{equation}
dc7_{unsrt} = [r,s,t,s+t-u,r+t-v,r+s-w,r+s+t-u-v-w+2\,min(u,v,w)]
  \label{ppp}
\end{equation}

If we subtract the face diagonal from the matching pairs of edges, we get the
absolute values of $u,v,w$
\[
dc7_{unsrt,2}+dc7_{unsrt,3}-dc7_{unsrt,4} 
\]
\[
= s+t-(s+t-|u|)=|u|
\]
\[
dc7_{unsrt,1}+dc7_{unsrt,3}-dc7_{unsrt,5}
\]
\[
= r+t-(r+t-|v|)=|v|
\]
\[
dc7_{unsrt,1}+dc7_{unsrt,2}-dc7_{unsrt,6}
\]
\[
= r+s-(r+s-|w|)=|w|
\]

\noindent{}from which we can compute an estimate, $\tau$, of the shortest
body diagonal that is exact for $---$ and a strict underestimate
for $+++$:
\[\tau\,= \,r+s+t-|u|-|v|-|w|\]
\[=\,dc7_{unsrt,1}\,+\,dc7_{unsrt,2}\,+\,dc7_{unsrt,3} \]
\[ -(dc7_{unsrt,2},+\,dc7_{unsrt,3},-\,dc7_{unsrt,4})\]
\[ -(dc7_{unsrt,1}\,+\,dc7_{unsrt,3}\,-\,dc7_{unsrt,5})\]
\[ -(dc7_{unsrt,1}\,+,dc7_{unsrt,2}\,-\,dc7_{unsrt,6})\]
\[=\,-\,dc7_{unsrt,1}\,-\,dc7_{unsrt,2}\,-\,dc7_{unsrt,3}\]
\[\,+\,dc7_{unsrt,4}\,+\,dc7_{unsrt,5}\,+ \,dc7_{unsrt,6}\]
  If $\tau=dc7_{unsrt,7}$,
we can be certain that all of $u,\,v,\,w\, \leq 0$.  If  $\tau \neq dc7_{unsrt,7}$ and the difference is larger than the possible experimental
or rounding errors, we can be certain that all of $u,v,w > 0$.  Thus the
Niggli cell can be recovered from $dc7_{unsrt}$.
Q.E.D.

\subsection{Recovering the components of the Bright example}

For example, the Niggli-reduced $\mathbf{G}^6$ versions of the cells in the
Bright example above are i: $ [ 6,\, 8,\, 10,\, 8,\, 4,\, 2] $ and ii: $ [ 6,\, 8,\, 10,\, -6,\, -2,\, -4 ]$.
Note that the former has all of $u,\, v,\, w$ positive and the latter 
has them all negative.
Table \ref{table::recover} shows the unsorted $\mathbf{DC}^7$ vectors
and the process of recovery of the $\mathbf{G}^6$ vectors.

\subsection{Examples using Phospholipase A2}

As noted in \cite{McGill2013}, the structures in the PDB for Phospholipase
A2 include slightly different experimental determinations that are presented
as different lattices, {\it e.g.} 1U4J in space group H3 as 
$ [ 80.36,\,	80.36,\,	99.44,\,	90,\,	90,\,	120 ] $,
 1G2X in space group C121 as
 
$ [ 80.949,\,	80.572,\,	57.098,\,	90,\,	90.35,\,	90] $ , and 1FE5 in space group R32 as 

$ [ 57.98,\,	57.98,\,	57.98,\,	92.02,\,	92.02,\,	92.02 ] $.

\noindent{}The primitive Niggli-reduced $\mathbf{G}^6$ $[r,s,t,u,v,w ]$ vectors are:

$ \text{1U4J:  } [ 3251.278,\,	3251.278,\,	3251.27,\,	44.826,\,	44.826,\,	44.826  ] $

$ \text{1G2X:  } [ 3260.182,\,	3261.147,\,	3261.147,\,	30.447,\,	28.234,\,	28.234  ] $

$ \text{1FE5:  } [ 3361.68,\,	3361.68,\,	3361.68,\,	-236.987,\,	-236.987,\,	-236.987 ] $

\noindent{}from which we compute the three $\mathbf{DC}^7$ vectors as

$ [ r,s,t,s+t-|u|, r+t-|v|, r+s-|w|, r+s+t-|u|-|v|-|w|+2 min(|u|,|v|,|w|)] $

\noindent{}in the first two $+++$ cases and as

$ [ r,s,t,s+t-|u|, r+t-|v|, r+s-|w|, r+s+t-|u|-|v|-|w|]$

\noindent{}in the final $---$ case:

$ \text{1U4J:  } [ 3251.278,\,	3251.278,\,	3251.278,\,	6457.73,\,	6457.73,\,	6457.73,\,	9709.008 ] $

$ \text{1G2X:  } [ 3260.182,\,	3261.147,\,	3261.147,\,	6491.847,\,	6493.095,\,	6493.095,\,	9752.029 ] $

$ \text{1FE5:  } [ 3361.68,\,	3361.68,\,	3361.68,\,	6486.373,\,	6486.373,\,	6486.373,\,	9374.079 ] $

In each case the original primitive Niggli-reduced $\mathbf{G}^6$ $[r,s,t,u,v,w ]$ vectors can be recovered by simply copying the first three elements of
the unsorted $\mathbf{DC}^7$ vector, $r,s,t$, then computing the three absolute values of $u,v,w$ as the differences $s+t$ minus the fourth element of the
unsorted $\mathbf{DC}^7$ vector, $r+t$ minus the fifth element of the
unsorted $\mathbf{DC}^7$ vector, and $r+s$ minus the sixth element.  
What remains is to compute $\tau = r+s+t-|u|-|v|-|w|$ and compare it to the
seventh element.  If they are the same to within rounding error, as they
are, indeed, for the 1FE5 case, then this is a $---$ case and the absolute
values of $u,v,w$ we have computed are the negatives of the actual values.
Otherwise this is a $+++$ case.  Completion of this example is left as an
exercise to the reader, but to help in checking your work, the three
values of $\tau$ are $9619.356$, $9695.561$, and $9374.079$. 
  
\subsection{Example of the effect of Small Perturbations an Face-centered Cell}

A well-know example occurs for face-centered cubic lattices. Choosing an initial face-centered (F-centered) unit cell of ($10 \sqrt(2)$, $10 \sqrt(2)$, $10 \sqrt(2)$, $90$, $90$, $90$), the primitive Niggli-reduced cell is ($10$, $10$, $10$, $60$, $60$, $60$). However, small changes in the original F-centered cell can change the 
primitive Niggli-reduced cell to ( $10....$, $10....$, $10....$, $90$, $120$, $120$ )
The cells  in Table \ref{table::fcent} are all derived from F-centered 
$ ( 14.14,\, 14.14,\, 14.14,\, 90,\, 120,\ 120 ) $ perturbing 
by 0.01\% normal to the vector.  The resulting unsorted $\mathbf{DC}^7$ vectors are given in Table \ref{table::fcentdc7}.

While these look very different, when the differences are smoothed
by permutations and gluing at boundaries, the distances among these cells are all small.

\section{The Boundaries of Unsorted $\mathbf{DC}^7$}

Whether we are working in seven dimensions with $\mathbf{DC}^7$ 
or in six dimensions with $\mathbf{G}^6$, the reduced cells
form a manifold for which it is useful to understand the
boundaries.   Inasmuch as seven-dimensional unsorted $\mathbf{DC}^7$ 
cells are invertibly derived from six-dimensional $\mathbf{G}^6$
cells, the six-dimensional boundary polytopes
of the manifold of valid Wigner-Seitz-reduced cells in $\mathbf{DC}^7$ can be derived directly
from the fifteen five-dimensional boundary polytopes of the manifold of valid Niggli-reduced
cells in $\mathbf{G}^6$ as described in \citeasnoun{Andrews2014} and then
applying equations (\ref{mmm}) and (\ref{ppp}) to the $\mathbf{G}^6$
descriptions of the boundaries.   The manifold
 of Wigner-Seitz-reduced cells in unsorted $\mathbf{DC}^7$
 is divided similarly to the way in which the manifold of Niggli-reduced cells is
 divided on the basis of whether $\tau < dc7_{usrt,7}$ for $+++$  or $\tau = dc7_{usrt,7}$
 for $---$.  The fifteen boundary polytopes are sufficient to then
 describe the primitive lattice characters, but, as with $\mathbf{G}^6$,
 seven additional special-position subspaces are needed to fully
 describe some of the centered cases.  See \citeasnoun{Andrews2014} for a
 discussion of the special-position subspaces.
 
 For consistency, we break from past tradition and use $u,v,w$ for
 both the all-positive and all-negative cases rather than using $-u,-v,-w$ for the
 $---$ case.
 
 Recall that in linear algebra an orthogonal projector $P$ into a subspace $S_{sub}$ of
 a space $S$ is a symmetric matrix that maps any element $s \in S$ into an element
 $s_{sub} \in S_{sub}$ and satisfies $P^2=P$.  Orthogonal projectors are
 commonly just called ``projectors''.  Note that $(I-P)s$ is orthogonal to
 $Ps$ for any projector $P$.  Projectors are often computed by {\it ad hoc} singular value decomposition (SVD).  In
 this case they were computed by the column-space operations of the
 William Schelter's symbolic algebra package GNU maxima \cite{christensen1994resources}, the open source version of DOE macsyma \cite{bogen1974macsyma}
 
 \subsection{Equal-cell-edge case}
 
 Recall that $r \leq s \leq t$ for Niggli-reduced cells.  The first two boundaries are the equal-edge boundary cases.  For both of these cases, the boundaries subdivide
 into one $---$ subcase and three $+++$ subcases, one for each of $u$, $v$, or $w$
 being minimal.  Then at least one of $r+s+t+u-v-w$, $r+s+t-u+v-w$, or $r+s+t-u-v+w$ respectively
 is the minimal body diagonal.
 
 \begin{itemize}
     \item{Case 1. $r=s$:  The cells in this case may be either $+++$ or $---$.\\
     $dc7_{unsrt,1}=dc7_{unsrt,2}$.
     
     $ dc7_{unsrt}  \text{ subspace 1}=[r,r,t,r+t-|u|,r+t-|v|,2r-|w|,$
     
     $ 2r+t-|u|-|v|-|w|+2 max(0,min(u,v,w))]$
     
     $ dc7_{unsrt} \text{ boundary 1 }---\text{ projector } =$
     
\begin{center}
\begin{equation*}
\begin{pmatrix}
\frac{5}{14}&\frac{5}{14}&\frac{-1}{7}&\frac{1}{7}
&\frac{1}{7}&\frac{1}{7}&\frac{-1}{7}\\[.25em]
\frac{5}{14}&\frac{5}{14}&\frac{-1}{7}&\frac{1}{7}
&\frac{1}{7}&\frac{1}{7}&\frac{-1}{7}\\[.25em]
\frac{-1}{7}&\frac{-1}{7}&\frac{6}{7}&\frac{1}{7}
&\frac{1}{7}&\frac{1}{7}&\frac{-1}{7}\\[.25em]
\frac{1}{7}&\frac{1}{7}&\frac{1}{7}&\frac{6}{7}&\frac{-1}{7}
&\frac{-1}{7}&\frac{1}{7}\\[.25em]
\frac{1}{7}&\frac{1}{7}&\frac{1}{7}&\frac{-1}{7}&\frac{6}{7}
&\frac{-1}{7}&\frac{1}{7}\\[.25em]
\frac{1}{7}&\frac{1}{7}&\frac{1}{7}&\frac{-1}{7}&\frac{-1}{7}
&\frac{6}{7}&\frac{1}{7}\\[.25em]
\frac{-1}{7}&\frac{-1}{7}&\frac{-1}{7}&\frac{1}{7}
&\frac{1}{7}&\frac{1}{7}&\frac{6}{7}
\end{pmatrix}
\end{equation*}
\end{center}

\[dc7_{unsrt} \text{ boundary 1 } +++ \text{ minimal }u \text{ projector } =\]
\begin{center}
\begin{equation*}
\begin{pmatrix}
\frac{1}{2}&\frac{1}{2}&0&0&0&0&0\\[.25em]
\frac{1}{2}&\frac{1}{2}&0&0&0&0&0\\[.25em]
0&0&\frac{4}{5}&\frac{1}{5}&\frac{-1}{5}&\frac{-1}{5}&\frac{1}{5}\\[.25em]
0&0&\frac{1}{5}&\frac{4}{5}&\frac{1}{5}&\frac{1}{5}&\frac{-1}{5}\\[.25em]
0&0&\frac{-1}{5}&\frac{1}{5}&\frac{4}{5}&\frac{-1}{5}&\frac{1}{5}\\[.25em]
0&0&\frac{-1}{5}&\frac{1}{5}&\frac{-1}{5}&\frac{4}{5}&\frac{1}{5}\\[.25em]
0&0&\frac{1}{5}&\frac{-1}{5}&\frac{1}{5}&\frac{1}{5}&\frac{4}{5}
\end{pmatrix}
\end{equation*}
\end{center}

\[dc7_{unsrt} \text{ boundary 1 } +++ \text{ minimal }v \text{ projector } =\]
\begin{center}
\begin{equation*}
\begin{pmatrix}
\frac{1}{2}&\frac{1}{2}&0&0&0&0&0\\[.25em]
\frac{1}{2}&\frac{1}{2}&0&0&0&0&0\\[.25em]
0&0&\frac{4}{5}&\frac{-1}{5}&\frac{1}{5}&\frac{-1}{5}&\frac{1}{5}\\[.25em]
0&0&\frac{-1}{5}&\frac{4}{5}&\frac{1}{5}&\frac{-1}{5}&\frac{1}{5}\\[.25em]
0&0&\frac{1}{5}&\frac{1}{5}&\frac{4}{5}&\frac{1}{5}&\frac{-1}{5}\\[.25em]
0&0&\frac{-1}{5}&\frac{-1}{5}&\frac{1}{5}&\frac{4}{5}&\frac{1}{5}\\[.25em]
0&0&\frac{1}{5}&\frac{1}{5}&\frac{-1}{5}&\frac{1}{5}&\frac{4}{5}
\end{pmatrix}
\end{equation*}
\end{center}

\[dc7_{unsrt} \text{ boundary 1 } +++ \text{ minimal }w \text{ projector } =\]
\begin{center}
\begin{equation*}
\begin{pmatrix}
\frac{5}{14}&\frac{5}{14}&\frac{1}{7}&\frac{-1}{7}&\frac{-1}{7}&\frac{1}{7}&\frac{1}{7}\\[.25em]
\frac{5}{14}&\frac{5}{14}&\frac{1}{7}&\frac{-1}{7}&\frac{-1}{7}&\frac{1}{7}&\frac{1}{7}\\[.25em]
\frac{1}{7}&\frac{1}{7}&\frac{6}{7}&\frac{1}{7}&\frac{1}{7}&\frac{-1}{7}&\frac{-1}{7}\\[.25em]
\frac{-1}{7}&\frac{-1}{7}&\frac{1}{7}&\frac{6}{7}&\frac{-1}{7}&\frac{1}{7}&\frac{1}{7}\\[.25em]
\frac{-1}{7}&\frac{-1}{7}&\frac{1}{7}&\frac{-1}{7}&\frac{6}{7}&\frac{1}{7}&\frac{1}{7}\\[.25em]
\frac{1}{7}&\frac{1}{7}&\frac{-1}{7}&\frac{1}{7}&\frac{1}{7}&\frac{6}{7}&\frac{-1}{7}\\[.25em]
\frac{1}{7}&\frac{1}{7}&\frac{-1}{7}&\frac{1}{7}&\frac{1}{7}&\frac{-1}{7}&\frac{6}{7}
\end{pmatrix}
\end{equation*}
\end{center}

\[dc7_{unsrt} \text{ boundary 1 transforms}=\]

$ ---: [s,r,t,v+t+r,u+t+s,w+s+r,w+v+u+t+s+r] $

$ +++: [s,r,t,-v+t+r,-u+t+s,-w+s+r,-w-v+2 min(u,v,w)-u+t+s+r] $

}
     
     \item{Case 2. $s=t$:  The cells in this case may be either $+++$ or $---$.\\
     $dc7_{unsrt,2}=dc7_{unsrt,3}$.
     
     $dc7_{unsrt}  \text{ subspace 2}=[r,s,s,2s-|u|,r+s-|v|,r+s-|w|,$
     
     $r+2s-|u|-|v|-|w|+min(max(0,u),max(0,v),max(0,w))]$
     
     $dc7_{unsrt} \text{ boundary 2 }---\text{ projector } =$
     
\begin{center}
\begin{equation*}
\begin{pmatrix}
\frac{6}{7}&\frac{-1}{7}&\frac{-1}{7}&\frac{1}{7}&\frac{1}{7}&\frac{1}{7}&\frac{-1}{7}\\[.25em]
\frac{-1}{7}&\frac{5}{14}&\frac{5}{14}&\frac{1}{7}&\frac{1}{7}&\frac{1}{7}&\frac{-1}{7}\\[.25em]
\frac{-1}{7}&\frac{5}{14}&\frac{5}{14}&\frac{1}{7}&\frac{1}{7}&\frac{1}{7}&\frac{-1}{7}\\[.25em]
\frac{1}{7}&\frac{1}{7}&\frac{1}{7}&\frac{6}{7}&\frac{-1}{7}&\frac{-1}{7}&\frac{1}{7}\\[.25em]
\frac{1}{7}&\frac{1}{7}&\frac{1}{7}&\frac{-1}{7}&\frac{6}{7}&\frac{-1}{7}&\frac{1}{7}\\[.25em]
\frac{1}{7}&\frac{1}{7}&\frac{1}{7}&\frac{-1}{7}&\frac{-1}{7}&\frac{6}{7}&\frac{1}{7}\\[.25em]
\frac{-1}{7}&\frac{-1}{7}&\frac{-1}{7}&\frac{1}{7}&\frac{1}{7}&\frac{1}{7}&\frac{6}{7}
\end{pmatrix}
\end{equation*}
\end{center}

\[dc7_{unsrt} \text{ boundary 2 } +++ \text{ minimal }u \text{ projector } =\]
\begin{center}
\begin{equation*}
\begin{pmatrix}
\frac{6}{7}&\frac{1}{7}&\frac{1}{7}&\frac{-1}{7}&\frac{1}{7}&\frac{1}{7}&\frac{-1}{7}\\[.25em]
\frac{1}{7}&\frac{5}{14}&\frac{5}{14}&\frac{1}{7}&\frac{-1}{7}&\frac{-1}{7}&\frac{1}{7}\\[.25em]
\frac{1}{7}&\frac{5}{14}&\frac{5}{14}&\frac{1}{7}&\frac{-1}{7}&\frac{-1}{7}&\frac{1}{7}\\[.25em]
\frac{-1}{7}&\frac{1}{7}&\frac{1}{7}&\frac{6}{7}&\frac{1}{7}&\frac{1}{7}&\frac{-1}{7}\\[.25em]
\frac{1}{7}&\frac{-1}{7}&\frac{-1}{7}&\frac{1}{7}&\frac{6}{7}&\frac{-1}{7}&\frac{1}{7}\\[.25em]
\frac{1}{7}&\frac{-1}{7}&\frac{-1}{7}&\frac{1}{7}&\frac{-1}{7}&\frac{6}{7}&\frac{1}{7}\\[.25em]
\frac{-1}{7}&\frac{1}{7}&\frac{1}{7}&\frac{-1}{7}&\frac{1}{7}&\frac{1}{7}&\frac{6}{7}
\end{pmatrix}
\end{equation*}
\end{center}

\[dc7_{unsrt} \text{ boundary 2 } +++ \text{ minimal }v \text{ projector } =\]
\begin{center}
\begin{equation*}
\begin{pmatrix}
\frac{4}{5}&0&0&\frac{-1}{5}&\frac{1}{5}&\frac{-1}{5}&\frac{1}{5}\\[.25em]
0&\frac{1}{2}&\frac{1}{2}&0&0&0&0\\[.25em]
0&\frac{1}{2}&\frac{1}{2}&0&0&0&0\\[.25em]
\frac{-1}{5}&0&0&\frac{4}{5}&\frac{1}{5}&\frac{-1}{5}&\frac{1}{5}\\[.25em]
\frac{1}{5}&0&0&\frac{1}{5}&\frac{4}{5}&\frac{1}{5}&\frac{-1}{5}\\[.25em]
\frac{-1}{5}&0&0&\frac{-1}{5}&\frac{1}{5}&\frac{4}{5}&\frac{1}{5}\\[.25em]
\frac{1}{5}&0&0&\frac{1}{5}&\frac{-1}{5}&\frac{1}{5}&\frac{4}{5}
\end{pmatrix}
\end{equation*}
\end{center}

\[dc7_{unsrt} \text{ boundary 2 } +++ \text{ minimal }w \text{ projector } =\]
\begin{center}
\begin{equation*}
\begin{pmatrix}
\frac{4}{5}&0&0&\frac{-1}{5}&\frac{-1}{5}&\frac{1}{5}&\frac{1}{5}\\[.25em]
0&\frac{1}{2}&\frac{1}{2}&0&0&0&0\\[.25em]
0&\frac{1}{2}&\frac{1}{2}&0&0&0&0\\[.25em]
\frac{-1}{5}&0&0&\frac{4}{5}&\frac{-1}{5}&\frac{1}{5}&\frac{1}{5}\\[.25em]
\frac{-1}{5}&0&0&\frac{-1}{5}&\frac{4}{5}&\frac{1}{5}&\frac{1}{5}\\[.25em]
\frac{1}{5}&0&0&\frac{1}{5}&\frac{1}{5}&\frac{4}{5}&\frac{-1}{5}\\[.25em]
\frac{1}{5}&0&0&\frac{1}{5}&\frac{1}{5}&\frac{-1}{5}&\frac{4}{5}
\end{pmatrix}
\end{equation*}
\end{center}

 \[dc7_{unsrt} \text{ boundary 2 transforms}=\]
 
 $ ---: [r,t,s,u+t+s,w+s+r,v+t+r,w+v+u+t+s+r] $
 
 $ +++: [r,t,s,-u+t+s,-w+s+r,-v+t+r,-w-v+2 min(u,v,w)-u+t+s+r]$
 
}
 \end{itemize}
 
 The special-position subspaces $\hat{1}$ and $\hat{2}$ are obtained by adding the constraints $1^\prime\!: \{u = v\}$ and
 $2^\prime\!:\{v=w\}$, respectively.

 \subsection{$90^\circ$ case}
 
 The $90^\circ$ case marks a possible transition between $---$ and $+++$.
 All the cells with a $90^\circ$ angle are in $---$.

 \begin{itemize}
     \item{Case 3. $u=0$:  The cells in this case must be $---$.\\
     $dc7_{unsrt,2}\,+\,dc7_{unsrt,3}\,-\,dc7_{unsrt,4}=0$.
     
     $dc7_{unsrt}  \text{ subspace 3}=[r,s,t,s+t,r+t+v,r+s+w,r+s+t+v+w]$
     
     $dc7_{unsrt} \text{ boundary 3 projector } =$
     
\begin{center}
\begin{equation*}
\begin{pmatrix}
\frac{3}{4}&0&0&0&\frac{1}{4}&\frac{1}{4}&\frac{-1}{4}\\[.25em]
0&\frac{2}{3}&\frac{-1}{3}&\frac{1}{3}&0&0&0\\[.25em]
0&\frac{-1}{3}&\frac{2}{3}&\frac{1}{3}&0&0&0\\[.25em]
0&\frac{1}{3}&\frac{1}{3}&\frac{2}{3}&0&0&0\\[.25em]
\frac{1}{4}&0&0&0&\frac{3}{4}&\frac{-1}{4}&\frac{1}{4}\\[.25em]
\frac{1}{4}&0&0&0&\frac{-1}{4}&\frac{3}{4}&\frac{1}{4}\\[.25em]
\frac{-1}{4}&0&0&0&\frac{1}{4}&\frac{1}{4}&\frac{3}{4}
\end{pmatrix}
\end{equation*}
\end{center}

     \[dc7_{unsrt} \text{ boundary 3 transform}=\]
     \[ [r,s,t,-u+t+s,v+t+r,w+s+r,w+v+2 min(u,-v,-w)-u+t+s+r]\]
     }
     \item{Case 4. $v=0$:  The cells in this case must be $---$.\\
     $dc7_{unsrt,1}\,+\,dc7_{unsrt,3}\,-\,dc7_{unsrt,5}=0$.
     
     $dc7_{unsrt} \text{ subspace 4}=[r,s,t,s+t+u,r+t,r+s+w, r+s+t+u+w]$
     
     $dc7_{unsrt} \text{ boundary 4 projector } =$ 
    
\begin{center}
\begin{equation*}
\begin{pmatrix}
\frac{2}{3}&0&\frac{-1}{3}&0&\frac{1}{3}&0&0\\[.25em]
0&\frac{3}{4}&0&\frac{1}{4}&0&\frac{1}{4}&\frac{-1}{4}\\[.25em]
\frac{-1}{3}&0&\frac{2}{3}&0&\frac{1}{3}&0&0\\[.25em]
0&\frac{1}{4}&0&\frac{3}{4}&0&\frac{-1}{4}&\frac{1}{4}\\[.25em]
\frac{1}{3}&0&\frac{1}{3}&0&\frac{2}{3}&0&0\\[.25em]
0&\frac{1}{4}&0&\frac{-1}{4}&0&\frac{3}{4}&\frac{1}{4}\\[.25em]
0&\frac{-1}{4}&0&\frac{1}{4}&0&\frac{1}{4}&\frac{3}{4}
\end{pmatrix}
\end{equation*}
\end{center}

      \[dc7_{unsrt} \text{ boundary 4 transform}=\]
      \[ [r,s,t,u+t+s,-v+t+r,w+s+r,w-v+u+2 min(-u,v,-w)+t+s+r]\]
      
     }
     \item{Case 5. $w=0$:  The cells in this case must be $---$.\\
     $dc7_{unsrt,1}\,+\,dc7_{unsrt,2}\,-\,dc7_{unsrt,6}=0$.
     
     $dc7_{unsrt} \text{ subspace 5}=[r,s,t,s+t+u,r+t+v,r+s,r+s+t+u+v]$
     
     $ dc7_{unsrt} \text{ boundary 5 projector } = $
\begin{center}
\begin{equation*}
\begin{pmatrix}
\frac{2}{3}&\frac{-1}{3}&0&0&0&\frac{1}{3}&0\\[.25em]
\frac{-1}{3}&\frac{2}{3}&0&0&0&\frac{1}{3}&0\\[.25em]
0&0&\frac{3}{4}&\frac{1}{4}&\frac{1}{4}&0&\frac{-1}{4}\\[.25em]
0&0&\frac{1}{4}&\frac{3}{4}&\frac{-1}{4}&0&\frac{1}{4}\\[.25em]
0&0&\frac{1}{4}&\frac{-1}{4}&\frac{3}{4}&0&\frac{1}{4}\\[.25em]
\frac{1}{3}&\frac{1}{3}&0&0&0&\frac{2}{3}&0\\[.25em]
0&0&\frac{-1}{4}&\frac{1}{4}&\frac{1}{4}&0&\frac{3}{4}
\end{pmatrix}
\end{equation*}
\end{center}

      \[dc7_{unsrt} \text{ boundary 5 transform}=\]
      \[[r,s,t,u+t+s,v+t+r,-w+s+r,-w+v+u+2 min(-u,-v,w)+t+s+r]\]
     }
 \end{itemize}
 
 In each $90^\circ$ case, the special-position subspace consists of $\hat{3}, \hat{4}, \hat{5}\!: \{u = v = w = 0\}$,
{\it i.e.} the primitive orthorhombic case, and we take
$3^\prime\!: \{v = w = 0\}$, $4^\prime\!: \{u = w = 0\}$, and $5^\prime\!: \{u = v = 0\}$.

\subsection{Face-diagonal  case}
 
 Recall that $|u| \leq s$, $|v| \leq r$, and $|w| \leq r$.  Equality marks
 the transition from edges being smaller than face diagonals to face diagonals
 possibly being smaller than the Niggli-reduced cell edges.

 \begin{itemize}
     \item{Case 6. $s=u, v \ge w$:  The cells in this case must be $+++$.\\
     $\tau=-\sum_{i=1}^3(-dc7_{unsrt,i})+\sum_{i=4}^6(dc7_{unsrt,i}) < dc7_{unsrt,7}$\\
     $dc7_{unsrt,2}=dc7_{unsrt,2}\,+\,dc7_{unsrt,3}\,-\,dc7_{unsrt,4}$\\
          \indent{}~~~~equivalent to
     $dc7_{unsrt,3}\,=\,dc7_{unsrt,4}$\\
     $dc7_{unsrt,3}\,-\,dc7_{unsrt,5} \ge
     dc7_{unsrt,2}\,-\,dc7_{unsrt,6}$.
     
     $dc7_{unsrt} \text{ subspace 6}=[r,s,t,t,r+t-v,r+s-w, r+t-v+w],\, v \geq w >0 $
     
     $dc7_{unsrt} \text{ boundary 6 projector } =$
     
\begin{center}
\begin{equation*}
\begin{pmatrix}
\frac{4}{5}&\frac{-1}{5}&0&0&\frac{-1}{5}&\frac{1}{5}&\frac{1}{5}\\[.25em]
\frac{-1}{5}&\frac{4}{5}&0&0&\frac{-1}{5}&\frac{1}{5}&\frac{1}{5}\\[.25em]
0&0&\frac{1}{2}&\frac{1}{2}&0&0&0\\[.25em]
0&0&\frac{1}{2}&\frac{1}{2}&0&0&0\\[.25em]
\frac{-1}{5}&\frac{-1}{5}&0&0&\frac{4}{5}&\frac{1}{5}&\frac{1}{5}\\[.25em]
\frac{1}{5}&\frac{1}{5}&0&0&\frac{1}{5}&\frac{4}{5}&\frac{-1}{5}\\[.25em]
\frac{1}{5}&\frac{1}{5}&0&0&\frac{1}{5}&\frac{-1}{5}&\frac{4}{5})
\end{pmatrix}
\end{equation*}
\end{center}

    \[dc7_{unsrt} \text{ boundary 6 transform}=\]
    \[ [r,s,-u+t+s,t,w-v-u+t+s+r,-w+s+r,-v+t+r] \]
     }
     
     \item{Case 7. $s=u, v < w$:  The cells in this case must be $+++$.\\
     $\tau=-\sum_{i=1}^3(-dc7_{unsrt,i})+\sum_{i=4}^6(dc7_{unsrt,i}) < dc7_{unsrt,7}$\\
     $dc7_{unsrt,2}=dc7_{unsrt,2}\,+\,dc7_{unsrt,3}\,-\,dc7_{unsrt,4}$\\
          \indent{}~~~~equivalent to
     $dc7_{unsrt,3}\,=\,dc7_{unsrt,4}$\\
     $dc7_{unsrt,3}\,-\,dc7_{unsrt,5} <
     dc7_{unsrt,2}\,-\,dc7_{unsrt,6}$.
     
     $ dc7_{unsrt} \text{ subspace 7}=[r,s,t,t,r+t-v,r+s-w, r+t+v-w],\, w > v > 0 $
     
     $ dc7_{unsrt} \text{ boundary 7 projector } = $
\begin{center}
\begin{equation*}
\begin{pmatrix}
\frac{6}{7}&\frac{1}{7}&\frac{-1}{7}&\frac{-1}{7}&\frac{1}{7}&\frac{-1}{7}&\frac{1}{7}\\[.25em]
\frac{1}{7}&\frac{6}{7}&\frac{1}{7}&\frac{1}{7}&\frac{-1}{7}&\frac{1}{7}&\frac{-1}{7}\\[.25em]
\frac{-1}{7}&\frac{1}{7}&\frac{5}{14}&\frac{5}{14}&\frac{1}{7}&\frac{-1}{7}&\frac{1}{7}\\[.25em]
\frac{-1}{7}&\frac{1}{7}&\frac{5}{14}&\frac{5}{14}&\frac{1}{7}&\frac{-1}{7}&\frac{1}{7}\\[.25em]
\frac{1}{7}&\frac{-1}{7}&\frac{1}{7}&\frac{1}{7}&\frac{6}{7}&\frac{1}{7}&\frac{-1}{7}\\[.25em]
\frac{-1}{7}&\frac{1}{7}&\frac{-1}{7}&\frac{-1}{7}&\frac{1}{7}&\frac{6}{7}&\frac{1}{7}\\[.25em]
\frac{1}{7}&\frac{-1}{7}&\frac{1}{7}&\frac{1}{7}&\frac{-1}{7}&\frac{1}{7}&\frac{6}{7}
\end{pmatrix}
\end{equation*}
\end{center}

 \[dc7_{unsrt} \text{ boundary 7 transform}=\]
 \[ [r,s,-u+t+s,t,-w+v-u+t+s+r,-w+s+r,-2 w+v+2 min(2 s-u,w,w-v)+t+r] \]
  }
     
     \item{Case 8. $s=-u$:  The cells in this case must be $---$.\\
    $\tau=-\sum_{i=1}^3(-dc7_{unsrt,i})+\sum_{i=4}^6(dc7_{unsrt,i}) = dc7_{unsrt,7}$\\
     $dc7_{unsrt,2}=dc7_{unsrt,2}\,\,+dc7_{unsrt,3}\,-\,dc7_{unsrt,4}$\\
     \indent{}~~~~equivalent to
     $dc7_{unsrt,3}\,=\,dc7_{unsrt,4}$.
     
     $ dc7_{unsrt} \text{ subspace 8}=[r,s,t,t,r+t+v,r+s+w,r+t+v+w],\, v, w \leq 0 $
     
     $ dc7_{unsrt} \text{ boundary 8 projector } =$
\begin{center}
\begin{equation*}
\begin{pmatrix}
\frac{4}{5}&\frac{-1}{5}&0&0&\frac{1}{5}&\frac{1}{5}&\frac{-1}{5}\\[.25em]
\frac{-1}{5}&\frac{4}{5}&0&0&\frac{1}{5}&\frac{1}{5}&\frac{-1}{5}\\[.25em]
0&0&\frac{1}{2}&\frac{1}{2}&0&0&0\\[.25em]
0&0&\frac{1}{2}&\frac{1}{2}&0&0&0\\[.25em]
\frac{1}{5}&\frac{1}{5}&0&0&\frac{4}{5}&\frac{-1}{5}&\frac{1}{5}\\[.25em]
\frac{1}{5}&\frac{1}{5}&0&0&\frac{-1}{5}&\frac{4}{5}&\frac{1}{5}\\[.25em]
\frac{-1}{5}&\frac{-1}{5}&0&0&\frac{1}{5}&\frac{1}{5}&\frac{4}{5}
\end{pmatrix}
\end{equation*}
\end{center}

      \[dc7_{unsrt} \text{ boundary 8 transform}=\]
      \[ [r,s,u+t+s,t,w+v+u+t+s+r,w+s+r,2 w+v+2 min(u+2s,-w-v,-w)+t+r]\]
     }

     \item{Case 9. $r=v, u \ge w$:  The cells in this case must be $+++$. \\
     $\tau=-\sum_{i=1}^3(-dc7_{unsrt,i})+\sum_{i=4}^6(dc7_{unsrt,i}) < dc7_{unsrt,7}$\\
     $dc7_{unsrt,1}=dc7_{unsrt,1}\,+\,dc7_{unsrt,3}\,-\,dc7_{unsrt,5}$\\
          \indent{}~~~~equivalent to
     $dc7_{unsrt,3}\,=\,dc7_{unsrt,5}$\\
     $dc7_{unsrt,3}\,-\,dc7_{unsrt,4} \ge
     dc7_{unsrt,1}\,-\,dc7_{unsrt,6}$.
     
     $ dc7_{unsrt} \text{ subspace 9}=[r,s,t,s+t-u,t,r+s-w,s+t-u+w],\, 0 < w \leq u $
     
     $ dc7_{unsrt} \text{ boundary 9 projector } =$
     
\begin{center}
\begin{equation*}
\begin{pmatrix}
\frac{4}{5}&\frac{-1}{5}&0&\frac{-1}{5}&0&\frac{1}{5}&\frac{1}{5}\\[.25em]
\frac{-1}{5}&\frac{4}{5}&0&\frac{-1}{5}&0&\frac{1}{5}&\frac{1}{5}\\[.25em]
0&0&\frac{1}{2}&0&\frac{1}{2}&0&0\\[.25em]
\frac{-1}{5}&\frac{-1}{5}&0&\frac{4}{5}&0&\frac{1}{5}&\frac{1}{5}\\[.25em]
0&0&\frac{1}{2}&0&\frac{1}{2}&0&0\\[.25em]
\frac{1}{5}&\frac{1}{5}&0&\frac{1}{5}&0&\frac{4}{5}&\frac{-1}{5}\\[.25em]
\frac{1}{5}&\frac{1}{5}&0&\frac{1}{5}&0&\frac{-1}{5}&\frac{4}{5}
\end{pmatrix}
\end{equation*}
\end{center}

    \[dc7_{unsrt} \text{ boundary 9 transform}=\]
    \[ [r,s,-v+t+r,w-v-u+t+s+r,t,-w+s+r,-u+t+s]\]
     }
     
     \item{Case A. $r=v, u < w$:  The cells in this case must be $+++$.\\
     $\tau=-\sum_{i=1}^3(-dc7_{unsrt,i})+\sum_{i=4}^6(dc7_{unsrt,i}) < dc7_{unsrt,7}$\\
     $dc7_{unsrt,1}=dc7_{unsrt,1}\,+\,dc7_{unsrt,3}\,-\,dc7_{unsrt,5}$\\
          \indent{}~~~~equivalent to
     $dc7_{unsrt,3}\,=\,dc7_{unsrt,5}$\\
     $dc7_{unsrt,3}\,-\,dc7_{unsrt,4} <
     dc7_{unsrt,1}\,-\,dc7_{unsrt,6}$.
     
     $ dc7_{unsrt} \text{ subspace A}=[r,s,t,s+t-u,t,r+s-w,s+t+u-w],\, 0 < u < w $
     
     $ dc7_{unsrt} \text{ boundary A projector } =$
     
\begin{center}
\begin{equation*}
\begin{pmatrix}
\frac{6}{7}&\frac{1}{7}&\frac{1}{7}&\frac{-1}{7}&\frac{1}{7}&\frac{1}{7}&\frac{-1}{7}\\[.25em]
\frac{1}{7}&\frac{6}{7}&\frac{-1}{7}&\frac{1}{7}&\frac{-1}{7}&\frac{-1}{7}&\frac{1}{7}\\[.25em]
\frac{1}{7}&\frac{-1}{7}&\frac{5}{14}&\frac{1}{7}&\frac{5}{14}&\frac{-1}{7}&\frac{1}{7}\\[.25em]
\frac{-1}{7}&\frac{1}{7}&\frac{1}{7}&\frac{6}{7}&\frac{1}{7}&\frac{1}{7}&\frac{-1}{7}\\[.25em]
\frac{1}{7}&\frac{-1}{7}&\frac{5}{14}&\frac{1}{7}&\frac{5}{14}&\frac{-1}{7}&\frac{1}{7}\\[.25em]
\frac{1}{7}&\frac{-1}{7}&\frac{-1}{7}&\frac{1}{7}&\frac{-1}{7}&\frac{6}{7}&\frac{1}{7}\\[.25em]
\frac{-1}{7}&\frac{1}{7}&\frac{1}{7}&\frac{-1}{7}&\frac{1}{7}&\frac{1}{7}&\frac{6}{7}
\end{pmatrix}
\end{equation*}
\end{center}

\[dc7_{unsrt} \text{ boundary A transform}=\]
\[[r,s,-v+t+r,-w-v+u+t+s+r,t,-w+s+r,-2 w+2 min(2r-v,w,w-u)+u+t+s]\]
     }
     
     \item{Case B. $r=-v$:  The cells in this case must be $---$.\\
     $\tau=-\sum_{i=1}^3(-dc7_{unsrt,i})+\sum_{i=4}^6(dc7_{unsrt,i}) = dc7_{unsrt,7}$\\
     $dc7_{unsrt,1}=dc7_{unsrt,1}\,+\,dc7_{unsrt,3}\,-\,dc7_{unsrt,5}$\\
          \indent{}~~~~equivalent to
     $dc7_{unsrt,3}\,=\,dc7_{unsrt,5}$.
     
     $ dc7_{unsrt} \text{ subspace B}=[r,s,t,s+t+u,t,r+s+w,s+t+u+w],\, u, w \leq 0 $
     
     $ dc7_{unsrt} \text{ boundary B projector } =$
     
\begin{center}
\begin{equation*}
\begin{pmatrix}
\frac{4}{5}&\frac{-1}{5}&0&\frac{1}{5}&0&\frac{1}{5}&\frac{-1}{5}\\[.25em]
\frac{-1}{5}&\frac{4}{5}&0&\frac{1}{5}&0&\frac{1}{5}&\frac{-1}{5}\\[.25em]
0&0&\frac{1}{2}&0&\frac{1}{2}&0&0\\[.25em]
\frac{1}{5}&\frac{1}{5}&0&\frac{4}{5}&0&\frac{-1}{5}&\frac{1}{5}\\[.25em]
0&0&\frac{1}{2}&0&\frac{1}{2}&0&0\\[.25em]
\frac{1}{5}&\frac{1}{5}&0&\frac{-1}{5}&0&\frac{4}{5}&\frac{1}{5}\\[.25em]
\frac{-1}{5}&\frac{-1}{5}&0&\frac{1}{5}&0&\frac{1}{5}&\frac{4}{5}
\end{pmatrix}
\end{equation*}
\end{center}

    \[dc7_{unsrt} \text{ boundary B transform}=\]
    \[[r,s,v+t+r,w+v+u+t+s+r,t,w+s+r,2 w+2 min(v+2 r,-w-u,-w)+u+t+s]\]
     }

     \item{Case C. $r=w, u \ge v$:  The cells in this case must be $+++$.\\
     $\tau=-\sum_{i=1}^3(-dc7_{unsrt,i})+\sum_{i=4}^6(dc7_{unsrt,i}) < dc7_{unsrt,7}$\\
     $dc7_{unsrt,1}=dc7_{unsrt,1}\,+\,dc7_{unsrt,2}\,-\,dc7_{unsrt,6}$\\
          \indent{}~~~~equivalent to
     $dc7_{unsrt,2}\,=\,dc7_{unsrt,6}$\\
     $dc7_{unsrt,2}\,-\,dc7_{unsrt,4} \ge
     dc7_{unsrt,1}\,-\,dc7_{unsrt,5}$.
     
     $ dc7_{unsrt} \text{ subspace C}=[r,s,t,s+t-u,r+t-v,s,s+t-u+v],\, 0 <v \leq u $
     
     $ dc7_{unsrt} \text{ boundary C projector } =$
     
\begin{center}
\begin{equation*}
\begin{pmatrix}
\frac{4}{5}&0&\frac{-1}{5}&\frac{-1}{5}&\frac{1}{5}&0&\frac{1}{5}\\[.25em]
0&\frac{1}{2}&0&0&0&\frac{1}{2}&0\\[.25em]
\frac{-1}{5}&0&\frac{4}{5}&\frac{-1}{5}&\frac{1}{5}&0&\frac{1}{5}\\[.25em]
\frac{-1}{5}&0&\frac{-1}{5}&\frac{4}{5}&\frac{1}{5}&0&\frac{1}{5}\\[.25em]
\frac{1}{5}&0&\frac{1}{5}&\frac{1}{5}&\frac{4}{5}&0&\frac{-1}{5}\\[.25em]
0&\frac{1}{2}&0&0&0&\frac{1}{2}&0\\[.25em]
\frac{1}{5}&0&\frac{1}{5}&\frac{1}{5}&\frac{-1}{5}&0&\frac{4}{5}
\end{pmatrix}
\end{equation*}
\end{center}

    \[dc7_{unsrt} \text{ boundary C transform}=\]
    \[[r,-w+s+r,t,-w+v-u+t+s+r,-v+t+r,s,-u+t+s]\]
     }
     \item{Case D. $r=w, u < v$:  The cells in this case must be $+++$.\\
     $\tau=-\sum_{i=1}^3(-dc7_{unsrt,i})+\sum_{i=4}^6(dc7_{unsrt,i}) <
     dc7_{unsrt,7}$\\
     $dc7_{unsrt,1}=dc7_{unsrt,1}\,+\,dc7_{unsrt,2}\,-\,dc7_{unsrt,6}$\\
          \indent{}~~~~equivalent to
     $dc7_{unsrt,2}\,=\,dc7_{unsrt,6}$\\
     $dc7_{unsrt,2}\,-\,dc7_{unsrt,4} <
     dc7_{unsrt,1}\,-\,dc7_{unsrt,5}$.
     
     $ dc7_{unsrt}\text{ subspace D} =[r,s,t,s+t-u,r+t-v,s,s+t+u-v],\, 0<u<v $
     $ dc7_{unsrt} \text{ boundary D projector } =$
     
\begin{center}
\begin{equation*}
\begin{pmatrix}
\frac{6}{7}&\frac{1}{7}&\frac{1}{7}&\frac{-1}{7}&\frac{1}{7}&\frac{1}{7}&\frac{-1}{7}\\[.25em]
\frac{1}{7}&\frac{5}{14}&\frac{-1}{7}&\frac{1}{7}&\frac{-1}{7}&\frac{5}{14}&\frac{1}{7}\\[.25em]
\frac{1}{7}&\frac{-1}{7}&\frac{6}{7}&\frac{1}{7}&\frac{-1}{7}&\frac{-1}{7}&\frac{1}{7}\\[.25em]
\frac{-1}{7}&\frac{1}{7}&\frac{1}{7}&\frac{6}{7}&\frac{1}{7}&\frac{1}{7}&\frac{-1}{7}\\[.25em]
\frac{1}{7}&\frac{-1}{7}&\frac{-1}{7}&\frac{1}{7}&\frac{6}{7}&\frac{-1}{7}&\frac{1}{7}\\[.25em]
\frac{1}{7}&\frac{5}{14}&\frac{-1}{7}&\frac{1}{7}&\frac{-1}{7}&\frac{5}{14}&\frac{1}{7}\\[.25em]
\frac{-1}{7}&\frac{1}{7}&\frac{1}{7}&\frac{-1}{7}&\frac{1}{7}&\frac{1}{7}&\frac{6}{7}
\end{pmatrix}
\end{equation*}
\end{center}

    \[dc7_{unsrt} \text{ boundary D transform}=\]
    \[[r,-w+s+r,t,-w-v+u+t+s+r,-v+t+r,s,2min(v,v-u,2r-w)-2v+u+t+s]\]

 }
     \item{Case E. $r=-w$:  The cells in this case must be $---$.\\
     $\tau=-\sum_{i=1}^3(-dc7_{unsrt,i})+\sum_{i=4}^6(dc7_{unsrt,i}) = dc7_{unsrt,7}$\\
     $dc7_{unsrt,1}=dc7_{unsrt,1}\,+\,dc7_{unsrt,2}\,-\,dc7_{unsrt,6}$\\
          \indent{}~~~~equivalent to
     $dc7_{unsrt,2}\,=\,dc7_{unsrt,6}$.
     
     $ dc7_{unsrt} \text{ subspace E } =[r,s,t,s+t+u,r+t+v,s,s+t+u+v], u, v \leq 0 $
     
     $ dc7_{unsrt} \text{ boundary E projector } =$
     
\begin{center}
\begin{equation*}
\begin{pmatrix}
\frac{4}{5}&0&\frac{-1}{5}&\frac{1}{5}&\frac{1}{5}&0&\frac{-1}{5}\\[.25em]
0&\frac{1}{2}&0&0&0&\frac{1}{2}&0\\[.25em]
\frac{-1}{5}&0&\frac{4}{5}&\frac{1}{5}&\frac{1}{5}&0&\frac{-1}{5}\\[.25em]
\frac{1}{5}&0&\frac{1}{5}&\frac{4}{5}&\frac{-1}{5}&0&\frac{1}{5}\\[.25em]
\frac{1}{5}&0&\frac{1}{5}&\frac{-1}{5}&\frac{4}{5}&0&\frac{1}{5}\\[.25em]
0&\frac{1}{2}&0&0&0&\frac{1}{2}&0\\[.25em]
\frac{-1}{5}&0&\frac{-1}{5}&\frac{1}{5}&\frac{1}{5}&0&\frac{4}{5}
\end{pmatrix}
\end{equation*}
\end{center}

    $ dc7_{unsrt} \text{ boundary E transform}=$
    
    $ [r,w+s+r,t,w+v+u+t+s+r,v+t+r,s,2v+2min((-v)-u,-v,w+2 r)+u+t+s]$

     }
 \end{itemize}
 
 The special-position subspaces of the face-diagonal boundary polytopes 6, 8, 9, B, C and E are empty because such a
special position would require a common point in the all acute $+ + +$ and all obtuse $- - -$ cases, but they only meet
at the axial planes of the $u,v,w$ subspace, which are excluded from the all acute $+ + +$ cases.  For cases
7, A and D there are non-trivial special-position subspaces. 
An invariant point in case 7 would have to satisfy $v = w - v$ or $v = w/2$.
Thus we define $7^\prime\!: \{v = w/2\}$ and similarly define $A^\prime\!: \{u = w/2\}$ and $D^\prime\!: \{u = v/2\}$.

\subsection{Body-diagonal case}

Recall that $t \leq r+s+t+u+v+w$ for a Niggli-reduced cell; otherwise
the main body diagonal would be shorter than $c$.  Equality can occur 
in $---$ and marks
the transition from edges being smaller than the main body diagonal to
the main body diagonal possibly being smaller.

\begin{itemize}
    \item{case F. $t = r+s+t+u+v+w$.  The cells in this case must be $---$.
    \[\tau=-\sum_{i=1}^3(-dc7_{unsrt,i})+\sum_{i=4}^6(dc7_{unsrt,i})\]
    \[=dc7_{unsrt,7}=dc7_{unsrt,3}\]
    
    $ dc7_{unsrt} \text{ subspace F}=[r,s,t,s+t+u,r+t+v,-u-v,t],\, u, v \leq 0 $
    
    $ dc7_{unsrt} \text{ boundary F projector}=$
\begin{center}
\begin{equation*}
\begin{pmatrix}
\frac{6}{7}&\frac{-1}{7}&\frac{-1}{7}&\frac{1}{7}&\frac{1}{7}&\frac{1}{7}&\frac{-1}{7}\\[.25em]
\frac{-1}{7}&\frac{6}{7}&\frac{-1}{7}&\frac{1}{7}&\frac{1}{7}&\frac{1}{7}&\frac{-1}{7}\\[.25em]
\frac{-1}{7}&\frac{-1}{7}&\frac{5}{14}&\frac{1}{7}&\frac{1}{7}&\frac{1}{7}&\frac{5}{14}\\[.25em]
\frac{1}{7}&\frac{1}{7}&\frac{1}{7}&\frac{6}{7}&\frac{-1}{7}&\frac{-1}{7}&\frac{1}{7}\\[.25em]
\frac{1}{7}&\frac{1}{7}&\frac{1}{7}&\frac{-1}{7}&\frac{6}{7}&\frac{-1}{7}&\frac{1}{7}\\[.25em]
\frac{1}{7}&\frac{1}{7}&\frac{1}{7}&\frac{-1}{7}&\frac{-1}{7}&\frac{6}{7}&\frac{1}{7}\\[.25em]
\frac{-1}{7}&\frac{-1}{7}&\frac{5}{14}&\frac{1}{7}&\frac{1}{7}&\frac{1}{7}&\frac{5}{14}
\end{pmatrix}
\end{equation*}
\end{center}

    \[dc7_{unsrt} \text{ boundary F transform}=\]
    \[[r,s,w+v+u+t+s+r,v+t+r,u+t+s,w+s+r,t] \]
    }
\end{itemize}

In order to have a special-position subspace in case F, in addition to $r+ s + t  + u + v +  w = t$, we need $u = -2 s - u -w$ and $v = -2 r - v - w$. From this we have 
$2 s + 2 u = -w = 2 r + 2 v$. Then it follows that $F^\prime\!:\{r - s - u+v = 0\}$.  
This is equivalent to
\(\|\textbf{a}+\textbf{c}\| = \|\textbf{b}+\textbf{c}\|\), \textit{i.e.}
the shorter b-face-diagonal is the same length as the shorter  a-face-diagonal.

The unsorted $\mathbf{DC}^7$ subspace descriptions of the 
non-anorthic lattice characters derived from the $\mathbf{G}^6$
subspace descriptions given in Table 4 in \citeasnoun{Andrews2014} are
presented here in Tables \ref{NiggliFormsI} and \ref{NiggliFormsII}  by applying equations (\ref{mmm}) and (\ref{ppp}).  Except for three
of the monoclinic cases, 55A, 55B and 57C, the boundary polytopes from
$\mathbf{G}^6$ fully describe those for unsorted $\mathbf{DC}^7$.
In those three cases, the boundary is divided between 
$0 < u \leq r$ and $r < u \leq s$ with two different forms each for
$dc7_{unsrt,7}$, the minimum body diagonal.  Such divided boundaries due to overlapping inequalities are encountered in Niggli reduction even when
working just in $\mathbf{G}^6$, and, just as in those cases, this adds combinatorial complexity that needs
to be allowed for in distance calculations.

\section{Smoothing by permutations}

Because the same boundaries are available in unsorted $\mathbf{DC}^7$ as
in $\mathbf{G}^6$, the equivalent algorithmic techniques can be used
in improving the distance calculations to improve smoothness.  The obvious
first step is to deal with boundary cases 1 and 2 by simple permutation
of the $dc7_{unsrt}$ vectors, so that
\[
  dc7unsrt\_dist(dc7_{1},dc7_{2})=min(
\]
\[
  ||[dc7_{1,1},dc7_{1,2},dc7_{1,3},dc7_{1,4},dc7_{1,5},dc7_{1,6},dc7_{1,7}]
\]
\[
  -[dc7_{2,1},dc7_{2,2},dc7_{2,3},dc7_{2,4},dc7_{2,5},dc7_{2,6},dc7_{2,7}]||,
\]
\[
  ||[dc7_{1,1},dc7_{1,2},dc7_{1,2},dc7_{1,4},dc7_{1,5},dc7_{1,6},dc7_{1,7}]
\]
\[
  -[dc7_{2,1},dc7_{2,3},dc7_{2,2},dc7_{2,4},dc7_{2,6},dc7_{2,5},dc7_{2,7}]||,
\]
\[
  ||[dc7_{1,1},dc7_{1,2},dc7_{1,2},dc7_{1,4},dc7_{1,5},dc7_{1,6},dc7_{1,7}]
\]
\[
  -[dc7_{2,2},dc7_{2,1},dc7_{2,3},dc7_{2,5},dc7_{2,4},dc7_{2,6},dc7_{2,7}]||,
\]
\[
  ||[dc7_{1,1},dc7_{1,2},dc7_{1,2},dc7_{1,4},dc7_{1,5},dc7_{1,6},dc7_{1,7}]
\]
\[
  -[dc7_{2,2},dc7_{2,3},dc7_{2,1},dc7_{2,5},dc7_{2,6},dc7_{2,4},dc7_{2,7}]||,
\]
\[
  ||[dc7_{1,1},dc7_{1,2},dc7_{1,2},dc7_{1,4},dc7_{1,5},dc7_{1,6},dc7_{1,7}]
\]
\[
  -[dc7_{2,3},dc7_{2,1},dc7_{2,2},dc7_{2,6},dc7_{2,4},dc7_{2,5},dc7_{2,7}]||,
\]
\[
  ||[dc7_{1,1},dc7_{1,2},dc7_{1,2},dc7_{1,4},dc7_{1,5},dc7_{1,6},dc7_{1,7}]
\]
\[
  -[dc7_{2,3},dc7_{2,2},dc7_{2,1},dc7_{2,6},dc7_{2,5},dc7_{2,4},dc7_{2,7}]||.
\]

These cases are simple because cases 1 and 2 do not impact the seventh element.
In the general case, {\it e.g.} a boundary transform, rather than a
simple Niggli-reduced cell-edge permutation, a fresh Niggli reduction may be needed to regenerate the
seventh element for minimal distance calculations.

\section{Testing against the Gruber example}

\citeasnoun{Gruber1973} presented a Niggli-reduced cell with a five-fold
Buerger-reduced cell ambiguity.  The Niggli-reduced cell is
$[a, b, c, \alpha ,\beta ,\gamma ]$ 
= $ [2,\,4,\,4,\,60,\,79.19,\,75.52] $
which is equivalent to the $\mathbf{G}^6$ cell 
$ [r,s,t,u,v,w]$ = $ [4,\, 16,\, 16,\, 16,\, 3,\, 4 ] $
and the unsorted $\mathbf{DC}^7$ cell $[4,\, 16,\, 16,\, 17,\, 19,\, 16,\, 16 ]$.  The five examples of the alternative
Buerger reduced cells are shown in Table \ref{table::grubercells} as edges and angles, in Table \ref{table::gruberg6cells} as $\mathbf{G}^6$ $[r,s,t,u,v,w]$, and in
Table \ref{table::gruberdc7unsrt} as unsorted $\mathbf{DC}^7$.
Cell i is Niggli-reduced.   All of the cells are on the 2 boundary
with $s=t$ and can equally be presented with $s$ and $t$ interchanged and $v$ and $w$ interchanged.  Cells i and ii
are both $+++$ and on the 7 and C face-diagonal boundaries as well as being on the 2 boundary.  Cells iii, iv and v are all $---$ and on the F body-diagonal boundary as well as being on the
2 boundary and one other face-diagonal boundary.
Cell iii is on the 8 face-diagonal boundary and cells iv and v
are on the E face-diagonal boundary.  Niggli reduction will
transform all of these back to cell i and Niggli reduction is
the first step in computing unsorted $\mathbf{DC}^7$. 
In order to compute the cells in Table \ref{table::gruberdc7unsrt} the components at the face-diagonal
boundaries were reduced in magnitude by 0.01 to prevent the
Niggli reduction from changing all the examples to be identical.  The differences among the unsorted $\mathbf{DC}^7$ cells are consistent with the perturbation.

\section{Summary and Conclusions}

Starting from a Niggli-reduced cell, a crystallographic lattice may be characterized by seven parameters describing the Dirichlet cell: three edge lengths, the three shorter face diagonals and the shortest body diagonal, from which the Niggli-reduced cell may be recovered.  This unsorted $\mathbf{DC}^7$ lattice characterization avoids the low-symmetry ambiguities of sorted $\mathbf{DC}^7$ and
is worth further investigation as a possible alternative to
$\mathbf{S}^6$ for crystallographic databases and clustering.


\appendix



\ack{{\bf Acknowledgements}}
\\

Our thanks to Frances C. Bernstein for careful copy-editing and
helpful suggestions.   Our thanks to Elizabeth Kincaid for the timeline figure.

Work supported in part by
 US Dept.~of Energy, Office of Science, DOE Office of Biological and Environmental Research, grant KP1607011 and 
DOE Office of Basic Energy Sciences Program, contract DE-SC0012704,
US NIH National Institute of General Medical Sciences P30GM133893,
U.S. Department of Energy, Office of Science, Office of Workforce Development for Teachers and Scientists (WDTS) under the Science Undergraduate Laboratory Internships Program (SULI).



\bibliographystyle{iucr}
\bibliography{DC7unsrt.bib}




\onecolumn

\begin{table}
	\caption{Cells derived from F-centered $ [ 14.14,\, 14.14,\, 14.14,\, 90,\, 90,\ 90 ]$ perturbed 
		by 0.01\% normal to the respective Selling vector and scaling $a$ to 10.  These are the primitive Niggli-reduced
		cells. They are sorted with alpha increasing.\\
	~~
}
	\label{table::fcent}
	\begin{tabular}{l c c c c c c}
                P&  10.000&  10.000&  10.003&  60.010&  60.038&  89.977\\
                P&  10.000&  10.003&  10.008&  60.017&  89.961&  60.031\\
                P&  10.000&  10.006&  10.008&  60.025&  60.046&  60.101\\
                P&  10.000&  10.002&  10.009&  60.039&  60.057&  60.082\\
                P&  10.000&  10.002&  10.006&  60.041&  60.028&  60.102\\
                P&  10.000&  10.001&  10.004&  60.047&  60.050&  60.016\\
                P&  10.000&  10.003&  10.006&  60.049&  60.036&  60.025\\
                P&  10.000&  10.000&  10.008&  60.049&  60.051&  60.030\\
                P&  10.000&  10.001&  10.002&  60.049&  60.059&  60.047\\
                P&  10.000&  10.005&  10.008&  60.052&  60.048&  60.076\\
                P&  10.000&  10.002&  10.004&  89.997&  60.032&  60.026\\
                P&  10.000&  10.001&  10.008&  90.018&  119.955& 119.980\\
                P&  10.000&  10.001&  10.004&  90.019&  119.984&  119.941\\
                P&  10.000&  10.005&  10.006&  90.051&  119.942&  119.972\\
                P&  10.000&  10.006&  10.007&  90.074&  119.940&  119.951\\
                P&  10.000&  10.001&  10.003&  119.938&  119.984&  90.024\\
                P&  10.000&  10.005&  10.005&  119.968&  119.978&  90.011\\
                P&  10.000&  10.000&  10.006&  119.971&  90.032&  119.963\\
                P&  10.000&  10.009&  10.012&  119.981&  90.022&  119.949\\
                P&  10.000&  10.010&  10.011&  119.990&  90.007&  119.947\\
\end{tabular}
\end{table}

\begin{table}
	\caption{unsorted $\mathbf{DC}^7$ vectors from the cells in Table \ref{table::fcent}.\\
	~~}
	\label{table::fcentdc7}
	\begin{tabular}{c c c c c c c}
		$r$&$s$&$t$&$s+t-|u|$&$r+t-|v|$&$r+s-|w|$&$r+s+t$\\
		&&&&&&$-|u|-|v|-|w|$\\
		&&&&&&$+2 max(0,$\\
		&&&&&&$min(u,v,w))$\\
                100.000&  100.006&  100.054&  100.060&  100.141&  199.927&  100.147\\
                100.000&  100.056&  100.154&  100.157&  200.017&  100.123&  100.224\\
                100.000&  100.119&  100.164&  100.216&  100.221&  100.367&  100.273\\
                100.000&  100.044&  100.181&  100.230&  100.263&  100.269&  100.312\\
                100.000&  100.046&  100.119&  100.206&  100.145&  100.333&  100.232\\
                100.000&  100.016&  100.078&  100.189&  100.190&  100.057&  100.230\\
                100.000&  100.054&  100.118&  100.234&  100.167&  100.104&  100.271\\
                100.000&  100.000&  100.152&  100.224&  100.232&  100.090&  100.304\\
                100.000&  100.020&  100.036&  100.178&  100.197&  100.152&  100.309\\
                100.000&  100.108&  100.168&  100.295&  100.228&  100.285&  100.355\\
                100.000&  100.013&  100.160&  200.109&  100.217&  100.067&  100.221\\
                100.000&  100.031&  100.072&  200.092&  100.133&  100.094&  100.228\\
                100.000&  100.017&  100.088&  200.037&  100.091&  100.187&  100.211\\
                100.000&  100.107&  100.130&  200.058&  100.240&  100.137&  100.199\\
                100.000&  100.118&  100.143&  200.004&  100.252&  100.209&  100.203\\
                100.000&  100.014&  100.055&  100.221&  100.076&  199.930&  100.159\\
                100.000&  100.090&  100.108&  100.195&  100.120&  200.050&  100.167\\
                100.000&  100.000&  100.125&  100.150&  200.014&  100.112&  100.152\\
                100.000&  100.179&  100.242&  100.269&  200.167&  100.243&  100.258\\
                100.000&  100.202&  100.227&  100.246&  200.202&  100.263&  100.280\\
\end{tabular}
\end{table}
\begin{table}
\caption{Bright's $\mathbf{DC}^7$ ambiguous example, redone in unsorted $\mathbf{DC}^7$.
The  Niggli-reduced $\mathbf{G}^6$ vectors are $ [ 6,\, 8,\, 10,\, 8,\, 4,\, 2] $ and $ [ 6,\, 8,\, 10,\, -6,\, -2,\, -4 ] $.  The former is $+++$ and 
becomes $ [ 6,\, 8,\, 10,\, 10,\, 12,\, 12,\, 14 ] $ as unsorted $\mathbf{DC}^7$.
The latter is $---$ and becomes $ [ 6,\, 8,\, 10,\, 12,\, 14,\, 10,\, 12 ] $
as unsorted $\mathbf{DC}^7$.  When each is processed to recover $\mathbf{G}^6$
the magnitude of $\mathbf{r}+\mathbf{s}+\mathbf{t}-\mathbf{|u|}-\mathbf{|v|}-\mathbf{|w|}$ disagrees with the minimum body diagonal for the former and agrees for the latter,
giving the correct signs for full recovery of $\mathbf{G}^6$.}
\label{table::recover}
\begin{center}
\begin{tabular}{|l|ccc|ccc|}
\hline
$\mathbf{G}^6$:&
$\mathbf{r}$&$\mathbf{s}$&$\mathbf{t}$&
$\mathbf{u}$&$\mathbf{v}$&$\mathbf{w}$\\
\hline
~~i&6& 8& 10& 8& 4& 2\\
~~ii&6&8&10&-6&-2&-4\\
\hline
\end{tabular}
\end{center}
\begin{center}
\begin{tabular}{|l|ccc|ccc|c|}
\hline
$\mathbf{DC}^7$ unsrt:&
$\mathbf{r}$&$\mathbf{s}$&$\mathbf{t}$&
$\mathbf{s}+\mathbf{t}$&
$\mathbf{r}+\mathbf{t}$&
$\mathbf{r}+\mathbf{s}$&
min body diag\\
&&&&$-\mathbf{|u|}$&$-\mathbf{|v|}$&$-\mathbf{|w|}$&(MBD)\\
\hline
~~i&6&8&10&10&12&12&14\\
~~ii&6&8&10&12&14&10&12\\
\hline
\end{tabular}
\end{center}
\begin{center}
\begin{tabular}{|l|ccc|ccc|c|}
\hline
recover $\mathbf{G}^6$:&
$\mathbf{r}$&$\mathbf{s}$&$\mathbf{t}$&
$\mathbf{|u|}$&$\mathbf{|v|}$&$\mathbf{|w|}$&$\tau=\mathbf{r}+\mathbf{s}+\mathbf{t}$\\
&&&&&&&$-\mathbf{|u|}-\mathbf{|v|}-\mathbf{|w|}$\\
\hline
~~i&6&8&10&8&4&2&$\tau\, 10 \neq \text{MBD}\, 14$ (disagree $+++$)\\
~~ii&6&8&10&6&2&4&$\tau\, 12 = \text{MBD}\, 12$ (agree $---$)\\
\hline
\end{tabular}
\end{center}
\end{table}

\begin{table}
\caption{\citeasnoun{Gruber1973} example of five-fold alternative Buerger-reduced
cells for a lattice as $[a,b,c,\alpha,\beta,\gamma]$}.
\label{table::grubercells}
\begin{center}
\begin{tabular}{|l|cccccc|}
\hline
cell&$a$&$b$&$c$&$\alpha$&$\beta$&$\gamma$\\
\hline
i&2&4&4&60.00&79.19&75.52\\
ii&2&4&4&60.00&86.42&75.52\\
iii&2&4&4&120.00&93.58&100.80\\
iv&2&4&4&117.95&93.58&104.48\\
v&2&4&4&113.97&100.80&104.48\\
\hline
\end{tabular}
\end{center}
\end{table}

\begin{table}
\caption{\citeasnoun{Gruber1973} example of five-fold alternative Buerger-reduced
cells for a lattice as\\
$\mathbf{G}^6$ $[\textbf{r},\textbf{s},\textbf{t},\textbf{u},\textbf{v},\textbf{w}]$.}
\label{table::gruberg6cells}
\begin{center}
\begin{tabular}{|l|cccccc|c|}
\hline
\textbf{cell}&$\textbf{r}$&$\textbf{s}$&$\textbf{t}$&$\textbf{u}$&$\textbf{v}$&$\textbf{w}$&\textbf{boundary}\\
\hline
i&4&16&16&16&3&4&27C\\      
ii&4&16&16&16&1&4&27C\\      
iii&4&16&16&-16&-1&-3&2F8\\      
iv&4&16&16&-15&-1&-4&2FE\\    
v&4&16&16&-13&-3&-4&2FE\\      
\hline
\end{tabular}
\end{center}
\end{table}

\begin{table}
\caption{\citeasnoun{Gruber1973} example of five-fold alternative Buerger-reduced
cells for a lattice as unsorted $\mathbf{DC}^7$.  To avoid the case where all the cells reduce to the same Niggli-reduced cell,
a small perturbation of .01 was applied to entries $u$ or $w$ in
$ii$ -- $v$ so that the vector did not sit precisely on the relevant boundaries.}.
\label{table::gruberdc7unsrt}
\begin{center}
\begin{tabular}{|l|ccc|ccc|c|}
\hline
cell&$1$&$2$&$3$&$4$&$5$&$6$&$7$\\
\hline
i&4&16&16&17&19&16&16\\
ii&4&16&16&17.01&19&16&16.01\\
iii&4&16&16&16.01&19&17&16.01\\
iv&4&16&16&17&19&16.01&16.01\\
v&4&16&16&19&17&16.01&16.01\\
\hline
\end{tabular}
\end{center}
\end{table}


\begin{table}
\caption{Roof/Niggli symbol, International Tables (IT) lattice character, Bravais lattice type, unsorted ${\mathbf{DC}^{\mathbf{7}}}$ subspace, boundary polytope.
Note that the variables r, s and t are non-negative, and u, v and w may be
positive, negative or zero as constrained below.}
\begin{center}
\begin{tabular}{|c|c|c|c|c|}
\hline
{\bf Roof/}     &{\bf IT}&{\bf Bravais}&{\bf Unsorted ${\mathbf{DC}^{\mathbf{7}}}$}&{\bf Bound-}\\
{\bf Niggli}&{\bf Lattice}&{\bf Lattice}&{\bf Subspace}&{\bf ary}\\
{\bf Symbol}&{\bf Char}&{\bf Type}&                             & {\bf Polytope} \\
\hline
44A&3&${\bf cP}$&$(r,r,r,2r,2r,2r,3r)$&$12345\!=\!12\hat{3}\!=\!12\hat{4}\!=\!12\hat{5}$\\
\hline
44C&1&$cF$&$(r,r,r,r,r,r,2r)$&12679ACD\\
\hline
44B&5&$cI$&$(r,r,r,4 r/3,4 r/3, 4 r/3,r)$&$\text{12F2}^{\prime} \text{F}^{\prime} = 1\hat{2}\hat{\text{F}}$\\
\hline
45A&11&${\bf tP}$&$(r,r,t,r+t,r+t,2r,2r+t)$&$1345 = 1\hat{3} = 1\hat{4} = 1\hat{5}$\\
45B&21&${\bf tP}$&$(r,s,s,2s,r+s,r+s,r+2s$&$2345 = 2\hat{3} = 2\hat{4} = 2\hat{5}$\\
\hline
45D&6&$tI$&$(r,r,r,r-w/2,r-w/2,2r+w,r),$&\\
&&&$-r \leq w \leq 0$&$\text{12FF}^{\prime} = 12\hat{\text{F}}$\\
45D&7&$tI$&$[r,r,r,2 r+u,r-u/2,r-u/2,r],$&\\
&&&$-r \leq u \leq 0$&$\text{12F2}^{\prime} = 1{\hat{2}}\text{F}$\\
45C&15&$tI$&$(r,r,t,t,t,2r,t)$&158BF\\
45E&18&$tI$&$(r,s,s,-r/2+2s,s,s,-r/2+2s)$&$\text{2ADA}^{\prime}  = 2{\hat{A}}\text{D}$\\
\hline
48A&12&${\bf hP}$&$(r,r,t,r+t,r+t,r,r+t])$&134E\\
48B&22&${\bf hP}$&$(r,s,s,s,r+s,r+s,r+s)$&2458\\
\hline
49C&2&$hR$&$(r,r,r,2r-u,2r-u,2r-u,3r-u),$&\\
&&&$0 < u \leq r$&$121^\prime 2^\prime = \hat{1}\hat{2}$\\
49D&4&$hR$&$(r,r,r,2r+u,2r+u,2r+u,3r+3u),$&\\
&&&$-r \leq u \leq 0$&$121^\prime 2^\prime = \hat{1}\hat{2}$\\
49B&9&$hR$&$(r,r,t,t,t,r,r+t)$&1679ACD\\
49E&24&$hR$&$(r,s,s,s+r/3,s+r/3,s+r/3,s)$&$2F2^\prime \text{F}^\prime = \hat{2}\hat{\text{F}}$\\
\hline
50C&32&${\bf oP}$&$(r,s,t,s+t,r+t,r+s,r+s+t)$&$345 = \hat{3} = \hat{4} = \hat{5}$\\
\hline
50D&13&$oC$&$(r,r,t,r+t,r+t,2r+w,2r+t+w),$&\\
&&&$-r \leq w \leq 0$&134\\
50E&23&$oC$&$(r,s,s,u+2 s,s+r,s+r,u+2 s+r),$&\\
&&&$-s \leq u \leq 0$&245\\
50A&36&$oC$&$(r,s,t,s+t,t,r+s,s+t)$&35B\\
50B&38&$oC$&$(r,s,t,s+t,r+t,s,s+t)$&34E\\
50F&40&$oC$&$[r,s,t,t,r+t,r+s,r+t)$&458\\
\hline
51A&16&$oF$&$(r,r,s,r+s+u,r+s+u,-2u,s),$&\\
&&&$-r \leq u \leq 0$&$\text{1F1}^{\prime} = \hat{1}\text{F}$\\
51B&26&$oF$&$(r,s,t,-r/2+s+t,t,s,-r/2+s+t)$&$\text{ADA}^{\prime}  = \hat{A}\text{D}$\\
\hline
52A&8&$oI$&$(r,r,r,2r+u,2r+v,-u-v,r),$&\\
&&&$-r \leq u \leq 0, -r \leq v \leq 0$&12F\\
52B&19&$oI$&$ (r,s,s,2s-u,s,s,-r+2s+u),$&\\
&&&$0 < u \leq r $&\\
&&&$ (r,s,s,2s-u,s,s,r+2s-u),$&\\
&&&$r < u \leq s $&29C = 2AD\\
52C&42&$oI$&$r,s,t,t,t,r+s,t$&58BF\\
\hline
\end{tabular}
\end{center}
\label{NiggliFormsI}
\end{table}%

\begin{table}
\caption{Roof/Niggli symbol, International Tables (IT) lattice character, Bravais lattice type, unsorted ${\mathbf{DC}^{\mathbi{7}}}$ subspace, boundary polytope, continued.  Note that the variables r, s and t are non-negative, and u, v and w may be
positive, negative or zero as constrained below.}
\begin{center}
\begin{tabular}{|c|c|c|c|c|}
\hline
{\bf Roof/}     &{\bf IT}&{\bf Bravais}&{\bf Unsorted ${\mathbf{DC}^{\mathbf{7}}}$}&{\bf Bound-}\\
{\bf Niggli}&{\bf Lattice}&{\bf Lattice}&{\bf Subspace}&{\bf ary}\\
{\bf Symbol}&{\bf Char}&{\bf Type}&                             & {\bf Polytope} \\
\hline
53A&33&${\bf mP}$&$(r,s,t,t+s,v+t+r,s+r,v+t+s+r),$&\\
&&&$-r \leq v \leq 0$&35\\
53B&35&${\bf mP}$&$(r,s,t,u+t+s,t+r,s+r,u+t+s+r),$&\\
&&&$-r \leq  u \leq 0$&45\\
53C&34&${\bf mP}$&$(r,s,t,t+s,t+r,w+s+r,w+t+s+r)$,4&\\
&&&$-r \leq w \leq 0$&34\\
\hline
55A&10, 14&$mC$&$(r,r,t,t+r-u,t+r-u,2 r-w$&\\
&&&$,t+2r-w), 0 < u \leq w \leq r $&\\
&&&$(r,r,t,t+r-u,t+r-u,2 r-w,$&\\
&&&$t+2 r-2 u +w), 0 < w < u \leq r $&$\text{11}^{\prime} = \hat{1}$\\
57B&17&$mC$&$(r,r,t,u+t+r,v+t+r,-u-v,t),$&\\
&&&$-r \leq u \leq 0$&1F\\
55B&20, 25&$mC$&$(r,s,s,2s-u,s+r-v,s+r-v,$&\\
&&&$2s+r+u-2 v), -s \leq u \leq v \leq 0$&\\
&&&$(r,s,s,2s-u,s+r-v,s+r-v,$&\\
&&&$2s+r-u), -r \leq v < u \leq 0$&$\text{22}^{\prime} = \hat{2}$\\
57C&27&$mC$&$(r,s,s,2s-u,s,s,-r+2s+u)$&\\
&&&$0 < u \leq r$&\\
&&&$(r,s,s,2s-u,s,s,r+2s-u)$&\\
&&&$r < u \leq s$&9C = AD\\
56A&28&$mC$&$(r,s,t,t+s-u,t,s+r-2u,t+s-u)$&\\
&&&$0 < u \leq r$&$\text{AA}^{\prime} = \hat{A}$\\
56C&29&$mC$&$(r,s,t,t+s-u,t+r-2u,s,t+s-u)$&\\
&&&$0 < u \leq r$&$\text{DD}^{\prime} = \hat{\text{D}}$\\
56B&30&$mC$&$(r,s,t,t,t+r-v,s+r-2v,+t-v)+r)$&\\
&&&$0 < v \leq r$&$\text{77}^{\prime} =  \hat{7}$\\
54C&37&$mC$&$(r,s,t,t+s+u,t,s+r,t+s+u), -r \leq u \leq 0 $&5B\\
54A&39&$mC$&$(r,s,t,t+s+u,t+r,s,t+s+u), -r \leq u \leq 0$&4E\\
54B&41&$mC$&$(r,s,t,t,v+t+r,s+r,v+t+r),$&\\
&&&$-r \leq v \leq 0;$&58\\
57A&43&$mC$&$(r,s,t,t-w/2,t-w/2,w+s+r,t),$&\\
&&&$-r \leq w \leq 0$&$\text{FF}^{\prime} = \hat{\text{F}}$\\
\hline
\end{tabular}
\end{center}
\label{NiggliFormsII}
\end{table}%

\end{document}